\documentclass[prb,showpacs,aps,floatfix,superscriptaddress, 11pt]{elsarticle}
\usepackage{amsmath,amssymb,graphicx,bm,color,mathrsfs,verbatim}
\pdfoutput=1

\begin{document}

\begin{frontmatter}
 \title{Many-body energy localization transition in periodically driven systems}
 \author[BU,KITP]{Luca D'Alessio\corref{cor1}}
 \author[BU]{Anatoli Polkovnikov\corref{cor2}}
 \address[BU]{Physics Department, Boston University, Boston, MA 02215, USA}
 \address[KITP]{Kavli Institute for Theoretical Physics, University of California, Santa Barbara, CA 93106, USA}
 \cortext[cor1]{E-mail address: dalessio@buphy.bu.edu}
 \cortext[cor2]{E-mail address: asp@bu.edu}

\begin{abstract}
According to the second law of thermodynamics the total entropy of a system is increased during almost any dynamical process. The positivity of the specific heat implies that the entropy increase is associated with heating. This is generally true both at the single particle level, like in the Fermi acceleration mechanism of charged particles reflected by magnetic mirrors, and for complex systems in everyday devices. Notable exceptions are known in noninteracting systems of particles moving in periodic potentials. Here the phenomenon of dynamical localization can prevent heating beyond certain threshold. The dynamical localization is known to occur both at classical (Fermi-Ulam model) and at quantum levels (kicked rotor). However, it was believed that driven ergodic systems will always heat without bound. Here, on the contrary, we report strong evidence of dynamical localization transition in both classical and quantum periodically driven ergodic systems in the thermodynamic limit. This phenomenon is reminiscent of many-body localization in energy space. 
\end{abstract}

\end{frontmatter}
\tableofcontents

\section{Introduction}
By stirring a cup of water its temperature is increased. When kneading a bread dough it gets warmer. These basic facts are common in our daily life and are a simple manifestation of the second law of thermodynamics at work. In fact, the second law of thermodynamics states that for almost any dynamical process (stirring, kneading, etc.) the entropy of the system needs to increase. The increase in entropy is usually associated with heating which we experience through our senses. The second law of thermodynamics can be rigorously proven microscopically if the initial state of the system is stationary, i.e. the initial density matrix is diagonal in the energy basis, and the probability distribution of occupying different energy states is a decreasing function of energy (so called passive density matrices)~\cite{thirring_02}. Then, as a result of any dynamical process, the energy of the system can only increase or stay the same. This is exactly the Thompson's formulation of the second law of thermodynamics~\cite{reif_09}. Similarly, with a single assumption of the initial state being stationary, one can prove that the properly defined entropy increases or stays constant~\cite{ap_ent}. Even more generally without any assumption on the initial state, it can be proven that the entropy and energy (for positive temperature states) increase in time for systems subject to random forces.  These systems include the famous Fermi acceleration problem, where a charged particle increases its energy by being repeatedly reflected by a magnetic mirror~\cite{fermi_49}, the Boltzmann's dynamics of particles experiencing repeated and uncorrelated collisions~\cite{LLX}, particles moving in a chaotic time-dependent cavity~\cite{jarzynski_93} and many others. In all previous examples it was assumed that there were no correlations between subsequent collisions. In quantum language this lack of correlations is equivalent to relaxation to the diagonal ensemble~\cite{srednicki_99, rigol_08} and it is sometimes referred to as ``energy measurements''~\cite{hanggi_11}. 

The applicability of the second law becomes much less clear if there are correlations between subsequent dynamical processes, like in the case of systems which are driven periodically in time. In these systems one can observe the phenomenon of dynamical localization, where the energy of the system never exceeds a maximum bound. Examples of dynamical localization include the Fermi-Ulam model of a particle bouncing off a periodically moving wall~\cite{lieberman_72}, classical and quantum kicked rotors~\cite{chirikov_81, fishman_82}, and the Kapitza pendulum~\cite{Kapitza,Kapitza-math}. It is generally expected that this localization phenomenon is peculiar to small integrable systems. In chaotic ergodic systems the periodicity of the driving should not matter because such systems effectively serve as their own heat bath, which in turn can be viewed as a source of a random Langevin type noise. Thus, it is generally expected that the energy of a periodically driven ergodic system will steadily increase in time. This belief though is entirely based on intuition and there is no guarantee that it is generally correct. Moreover, using the example of the many-body localization transition, it has been convincingly argued that disordered interacting systems can behave non-ergodically~\cite{basko_06, pal_10} and the standard assumption that the interacting non-integrable system can serve as its own bath can fail. Furthermore, for driven interacting systems it was recently noticed numerically that periodic modulation, even with a long period, leads to suppression of energy and entropy growth~\cite{ap_ent, Boris} compared to a a similar modulation with a random period. However, it was not clear whether this suppression simply leads to a slower heating rate or to the localization.

The purpose of this work is to argue that there is a new type of localization transition in periodically driven systems as a function of the driving period. We demonstrate this transition both analytically and numerically and argue that it is related to a breakdown of short time (Magnus) expansion of the evolution operator. For short periods the expansion is convergent leading to the effective time-independent many-body (Floquet) Hamiltonian and the energy is localized. For long periods this expansion breaks in the thermodynamic limit leading to the delocalization transition and heating of the system to the infinite temperature. These expectations are consistent with the recent experimental findings on a AC-driven electron glass \cite{Ovadyahu}. In this experiment, the energy absorbed by the electrons from the AC-driving, is related to the variation in the conductance which can be directly measured and it is convincingly shown that at high frequency (short period) the electron glass does not absorb energy. Moreover, it is shown that the critical frequency is set by the electron-phonon interactions and it is much lower than the maximum rate of energy exchange which is set by electron-electron interactions. Finally, we will show a strong evidence for this transition using examples of classical and quantum interacting spin systems.

\section{The Kapitza pendulum}

\begin{figure}
\includegraphics[width=1.0\columnwidth]{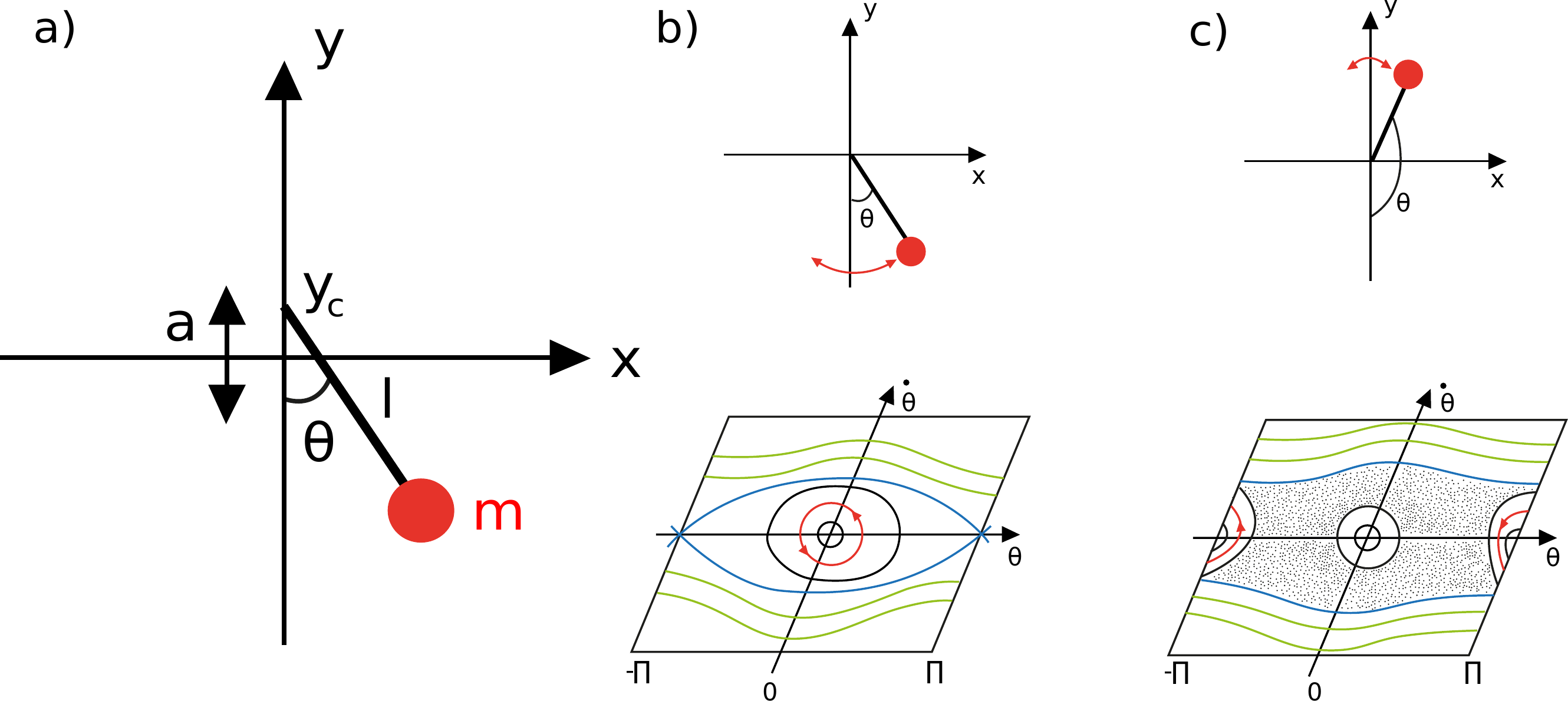}
\caption{ Schematic representation of the Kapitza pendulum, i.e. a rigid pendulum with vertically oscillating point of suspension, and its phase portraits. (a) The Kapitza pendulum. (b) Non-driven regime: the pendulum performs small oscillations around the stable lower equilibrium which are represented by the red line in the phase portrait.  (c) Dynamical stabilization regime: the pendulum performs small oscillations around the stable upper equilibrium which are represented by the red line in the phase portrait. In the phase portraits the green lines correspond to rotations, the black lines to oscillations, the blue lines are the separatrices and the points represent the region of chaotic motion. (For interpretation of the references to colour in this figure legend, the reader is referred to the web version of this article.)
}
\label{fig:kapitza}
\end{figure}

Before addressing the many-particle problem we will discuss a much simpler example of a periodically driven system, the so called Kapitza pendulum~\cite{Kapitza} and show how the Magnus expansion can be used to derive the effective potential. The Kapitza pendulum is a classical rigid pendulum with a vertically oscillating point of suspension (see Fig.~\ref{fig:kapitza}).

The equation of motion of the Kapitza pendulum reads
\begin{equation}
\ddot{\theta}=-\left(\omega_{0}^{2}+\frac{a}{l}\gamma^{2}\cos\left(\gamma t\right)\right)\sin\theta\label{eq:kapitza}
\end{equation}
where $\theta$ is the angle measured from the downward position (see Fig.~\ref{fig:kapitza}),
$\omega_{0}=\sqrt{\frac{g}{l}}$ is the frequency of small oscillations and $a,\,\gamma$ are the amplitude and frequency of the driving of
the point of suspension: $y_{c}=-a\,\cos\left(\gamma t\right)$. This
dynamical system has an extremely rich behavior containing both regions of chaotic and regular motion (see Ref.~\cite{Kapitza-math} and references therein).  For our purposes we consider the limit of small driving amplitude $a/l\ll 1$ and describe
how the dynamical behavior qualitatively changes \textit{as a function of the driving frequency}.

For a small amplitude drive the lower equilibrium at $\theta=0$ remains stable unless particular parametric resonance conditions $\frac{\gamma}{\omega_{0}}\approx\frac{2}{n}$ with $n=integer$, are met~\cite{Landau}. As we increase the frequency of the external drive 
from $\gamma\approx2\omega_{0}$ we observe qualitatively different regimes. First the motion in phase space is completely chaotic and both the lower and upper equilibrium ($\theta=0,\pi$) are unstable, then the lower equilibrium becomes stable while the upper equilibrium remains unstable, finally when $\frac{a}{l}\,\frac{\gamma}{\omega_{0}}>\sqrt{2}$ both the upper and lower equilibrium are stable. The surprising phenomenon that the upper position becomes stable (and the pendulum performs oscillations around this inverted position) is known in the literature as dynamical stabilization and was first explained by Kapitza. He showed that for small amplitude and high frequency driving the dynamic of the driven pendulum can be accurately described by a \textit{time-independent effective Hamiltonian}, moreover the effective potential energy develops a local minimum at $\theta=\pi$ when $\frac{a}{l}\,\frac{\gamma}{\omega_{0}}>\sqrt{2}$ explaining the oscillations around the inverted position. 

Usually the dynamical stabilization is obtained by splitting the degrees of freedom into fast and slow modes, eliminating the fast modes, and obtaining the effective potential for the slow modes~\cite{Landau}. This procedure has a limitation that it can not be easily extended to either interacting systems or to the quantum domain. It is also unclear whether averaging over fast degrees of freedom will lead to the Hamiltonian equation of motion in each order of the expansion. Here we show that the dynamical stabilization phenomenon can be understood through the Magnus expansion of the quantum evolution operator in powers of the inverse frequency (a relate perturbative analysis in powers of the inverse frequency was studied in \cite{fishman-1,fishman-2}) . The advantage of this method is that allows us to analyze behavior of the periodically driven interacting systems.

\subsection{Magnus Expansion}
Let us now start by reviewing the basis aspects of the Magnus Expansion~\cite{Magnus} (ME)
(for an excellent review see Ref.~\cite{Magnus-report}). This expansion equally applies to quantum and classical systems, but it is more convenient to use the language of quantum mechanics. We imagine that the system is prepared initially in some quantum-mechanical state $|\psi_0\rangle$, which evolves in time under some periodic Hamiltonian $H(t)$ with a period $T$. Then if we look stroboscopically into the system at times $t_n=nT$ its wave function will be given by
\begin{equation}
|\psi(nT)\rangle=U(T)^n |\psi_0\rangle
\end{equation}
where $U(T)$ is the evolution operator during one period of oscillation:
\begin{equation}
U(T)=T_\tau  \,\exp\left(-{i\over \hbar}\int_{0}^{T}d\tau H(t)\right).\label{eq:U1}
\end{equation}
Here the time ordering exponent implies that the later times in the integral always appear on the left. Since the evolution operator is unitary it can be always represented as the exponent of some Hermitian operator, which we can identify with the effective Floquet Hamiltonian $H_{eff}$.
\begin{equation}
U(T)=\exp\left(-{i\over \hbar}H_{eff} T\right)\label{eq:U2}
\end{equation}
The problem of finding the Floquet Hamiltonian is thus equivalent to the problem of evaluating the logarithm of a time-ordered exponential. For some simple noninteracting Hamiltonians it is actually possible to find the Floquet Hamiltonian explicitly and this was recently used to predict new topological states of light and atoms~\cite{lindner_11, kitagawa_12}. But in general it is a very hard problem with no closed form solution because the Hamiltonian at different times does not commute with itself. Unlike in statistical physics this problem remains equally hard even in the classical limit because of the singular factor $1/\hbar$ in the exponent.  

A possible root to compute the Floquet Hamiltonian is to expand the evolution operator in Taylor series in the period $T$ and then take the logarithm of this Taylor series. The corresponding ME~\cite{Magnus, Magnus-report} is guaranteed to converge for finite-dimensional Hamiltonians and sufficiently short periods. In thermodynamic limit the convergence of the ME can not be proven in general. This situation is very similar to the high-temperature expansion in statistical physics~\cite{kardar}. Usually the breakdown of the high temperature expansion is an indication of a phase transition. For simple few body systems there are many physical situations where the ME converges quickly to the exact Floquet Hamiltonian and only few terms are necessary to describe the dynamics. For this reason the ME found many applications in different areas of physics and mathematics (see the review~\cite{Magnus-report} and refs. therein). The first two terms in the ME read:
\begin{equation}
\begin{array}{c}
\hat{H}_{eff}^{(1)}=\frac{1}{T}\int_0^T dt\hat{H}(t_{1}),\quad 
\hat{H}_{eff}^{(2)}=\frac{1}{2T\left(i\hbar\right)}\iint_{0<t_2<t_2<T} dt_1 dt_2\left[\hat{H}(t_{1}),\hat{H}(t_{2})\right].
\end{array}\label{magnus}
\end{equation}
Higher order terms have a similar structure containing higher order commutators. They are multiplied by higher powers of the period and hence the period of the driving plays the role similar to the inverse temperature in the high temperature expansion. For classical systems the equivalent expansion can be obtained by substituting the commutators between the operators with the Poisson brackets: $[\dots]/{i\hbar} \to \{\dots\}$. 

In the case of the Kapitza pendulum it is sufficient to compute the first three terms in the ME (see Appendix \ref{appendix-kapitza}), i.e. $H_{eff}\approx H_{eff}^{(1)}+H_{eff}^{(2)}+H_{eff}^{(3)}$. They read:
\begin{equation}
\begin{array}{c}
H_{eff}^{(1)}=\frac{1}{2m}p^{2}-m\omega_{0}^{2}\cos\theta, \quad H_{eff}^{(2)}=0\\
H_{eff}^{(3)}=\frac{a}{l}{p^{2}\over m}\cos\theta+m\omega_0^2 \left(\left(\frac{\gamma a}{2 l \omega_{0}}\right)^2-\frac{a}{l}\right)\sin^{2}\theta
\end{array}\label{eq:Kapitza-classical-1}
\end{equation}
We point that up to this order, the ME is the same for quantum and classical pendulum. Here, the first term is simply the time averaged Hamiltonian, which reduces to the Hamiltonian of the non-driven pendulum. The second term is zero. Actually it can be shown that for symmetric protocol ($H(t)=H(T-t)$) all even terms in the ME are identically equal to zero~\cite{Magnus-report, even-order1, even-order2}. The third term $H_{eff}^{\left(3\right)}$ is the first nontrivial contribution to the Floquet Hamiltonian, which is sufficient to explain the dynamical stabilization of the pendulum found by Kapitza. Indeed in the limit $a\ll l$ the Floquet Hamiltonian describes a pendulum moving in the effective potential $U_{eff}=m\omega_{0}^{2}\left(-\cos\theta+\left(\frac{a\gamma}{2l\omega_{0}}\right)^{2}\sin^{2}\theta\right)$. It is easy to see that this potential develops a new minimum at the inverted position, $\theta=\pi$ when the condition $\frac{a}{l}\,\frac{\gamma}{\omega_{0}}> \sqrt{2}$ is met. Moreover one can check that the higher order terms in the ME remain small as long as this condition is met. When the period of the external driving is increased (the frequency is reduced) higher order terms in the ME need to be computed to accurately describe the dynamics of the system. If the period is increased even further there is no guarantee that the ME converges to the effective Floquet Hamiltonian. In this regime we observe that the motion of the pendulum becomes chaotic~\cite{Kapitza-math}. Recently the dynamical stabilization of an interacting quantum system, conceptually similar to the Kapitza's case described here, has been experimentally demonstrated \cite{chapman}. In this experiment, using a time-periodic protocol, the internal spin dynamics of a multi-component spinor condensate is stabilized and the region of stability computed using the time averaged Hamiltonian compares favorably with the experimental results. We suspect that the inclusion of higher order terms in the Magnus expansion could further improve the agreement between the theoretical calculations and the experimental results.

\section{Energy localization transition in interacting spin systems}

\begin{figure}
\includegraphics[width=1\columnwidth]{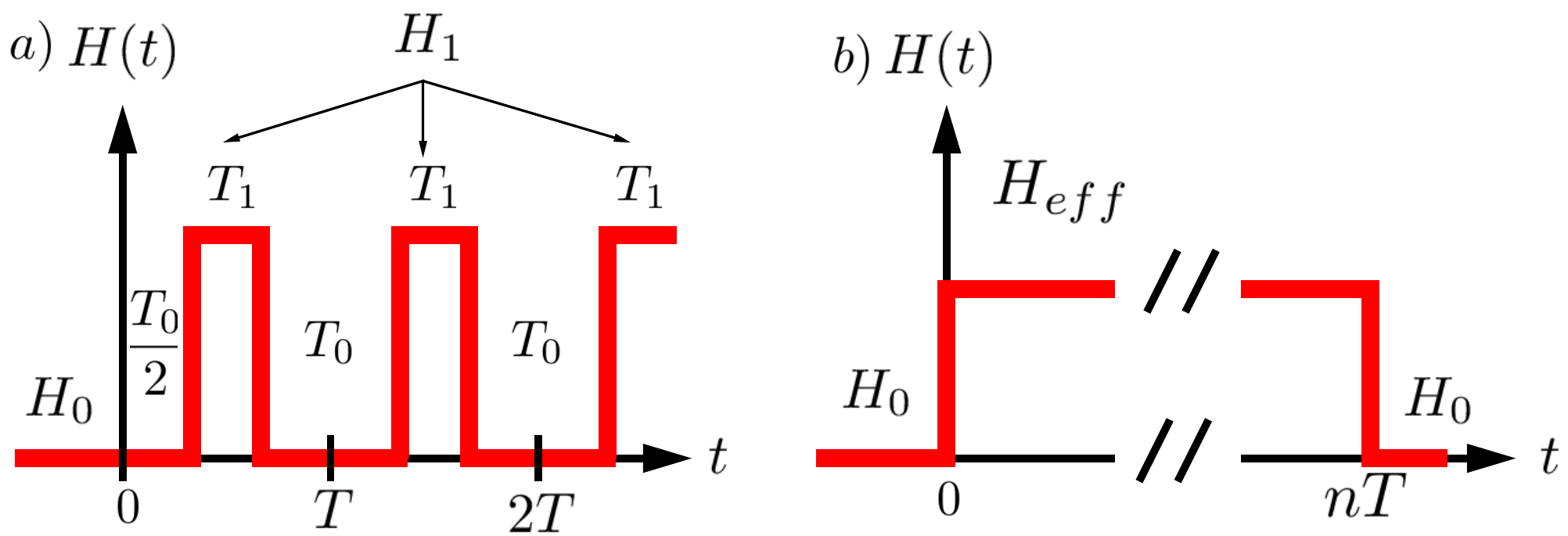}
\caption{Two equivalent description of the driving protocol:
(left) sequence of sudden quenches between $H_{0}$ and $H_{1}$ and
(right) single quench from $H_{0}$ to the effective Floquet Hamiltonian
$H_{eff}$ and back to $H_{0}$.\label{fig:protocol}}
\end{figure}

In the rest of the paper we will focus an $1d$ interacting classical and quantum spin models with periodic boundary conditions which is driven by periodically quenching between two different Hamiltonians, $H_{0}$ and $H_{1}$. Although we believe that our results are general and not limited $1d$ situations, we prefer to keep the discussion focused on this specific system.  In Fig. \ref{fig:protocol} we show two equivalent descriptions of the same protocol. The left panel represents the actual time dependent sequence of pulses switching between the two Hamiltonians. On the right panel we depict an equivalent protocol where instead we are performing a single quench to the Floquet Hamiltonian and then a quench back to the original Hamiltonian $H_0$ at the time of measurement. The second quench is not necessary, it highlights the fact that we do the measurements of observables like energy with respect to the Hamiltonian $H_0$.

Without loss of generality we can choose the Hamiltonian $H_0$ to be a simple precession in the external magnetic field and the Hamiltonian $H_1$ to be interacting and ergodic:
\begin{equation}
\begin{array}{c}
H_{0}=B_{x}H_{Bx},\quad 
H_{1}=J_{z}H_{z}+J_{z}^{\prime}H_{z}^{\prime}+J_{\parallel}H_{\parallel}+J_{\parallel}^{\prime}H_{\parallel}^{\prime}
\end{array}\label{eq:Hamiltonians}
\end{equation}
where, we have defined the shorthand notations::
\[
\begin{array}{l}
H_{Bx}=\sum_{n}\, s_{n}^{x},\quad 
H_{z}=\sum_{n}\,\left(s_{n}^{z}s_{n+1}^{z}\right),\,\, H_{\parallel}=\sum_{n}\,\left(s_{n}^{x}s_{n+1}^{x}+s_{n}^{y}s_{n+1}^{y}\right)\\
H_{z}^{\prime}=\sum_{n}\,\left(s_{n}^{z}s_{n+2}^{z}\right),\,\, H_{\parallel}^{\prime}=\sum_{i}\,\left(s_{n}^{x}s_{n+2}^{x}+s_{n}^{y}s_{n+2}^{y}\right)
\end{array}
\]
Let us point that this system is invariant under space translation and $\pi-rotation$ around the $x-axis$ ($s_{n}^{x}\rightarrow s_{n}^{x}$,
$s_{n}^{y}\rightarrow-s_{n}^{y}$, $s_{n}^{z}\rightarrow-s_{n}^{z}$). For numerical calculations we choose the following parameters: $B_{x}=1,\,\, J_{z}=-J_{\parallel}^{\prime}=\frac{1}{2},\,\, J_{z}^{\prime}=\frac{1}{40},\,\, J_{\parallel}=-\frac{1}{4}$.  We checked that our results are not tied to any particular choice of couplings.

As pointed out earlier, we can expect two qualitatively different regimes depending on the period of the driving. At long periods the system has enough time to relax to the stationary state between the pulses and thus is expected to constantly absorb energy until it reaches the infinite temperature. This situation is similar to what happens for driving with random periods~\cite{bunin_11}. On the contrary if the period is very short we can expect that the Floquet Hamiltonian converges to the time averaged Hamiltonian. Since the whole time evolution can be viewed as a single quench to the Floquet Hamiltonian (right panel in Fig.~\ref{fig:protocol}) we expect that the energy will be localized even in the infinite time limit as long as the Floquet Hamiltonian is well defined and local. Noticing that the commutator of two local extensive operators is local and extensive we see from Eq.~(\ref{magnus}) that the Floquet Hamiltonian is local and extensive in each order of ME and the dimensionless expansion parameter in the ME is a product of the period of the driving and the coupling constants (this is in analogy with the high temperature expansion in thermodynamics in which the inverse temperature takes the role of the period of the driving). Thus the question of whether the energy of the system is localized in the infinite time limit or reaches the maximum possible value is tied to the question of convergence of the ME. To the best of our knowledge there are no statements in the literature about this convergence in the thermodynamic limit. Therefore we will rely on the specific spin model to establish that for short driving periods the ME indeed converges and the energy of the system is localized while for longer periods this expansion diverges and the systems is heated towards infinite temperature.

For the quantum system we consider a spin-$\frac{1}{2}$ chain initially polarized along the $x$-axis, which is the ground state of $H_{0}$. We then compute the time evolution operator by exact diagonalization of the two Hamiltonians $H_{0}$ and $H_{1}$. Translational invariance of the system allows us to restrict the analysis to the zero momentum sector and analyze chains with the number of spins up to $N=17$, which contains $7712$ states. For the system of classical spins we numerically integrate the Bloch equations derived from the Hamiltonians~\eqref{eq:Hamiltonians} using the fourth-order Runge-Kutta algorithm. We took the initial conditions which maximally mimic the Wigner function of the quantum chain, namely  $s_{x}=\frac{1}{2}$, $s_{y}=\frac{1}{\sqrt{2}}\cos\alpha$ and $s_{z}=\frac{1}{\sqrt{2}}\sin\alpha$
where $0\le\alpha\le2\pi$ is a uniformly distributed random variable. The advantage of quantum systems is that it is straightforward to exactly obtain the infinite time limit of the evolution by projecting the initial state to the eigenstates of the effective Hamiltonian, while the advantage of the classical systems is that one can analyze significantly larger system sizes.

\subsection{Excess energy of the quantum spin chain}

\begin{figure}[ht]
\includegraphics[width=0.9\columnwidth]{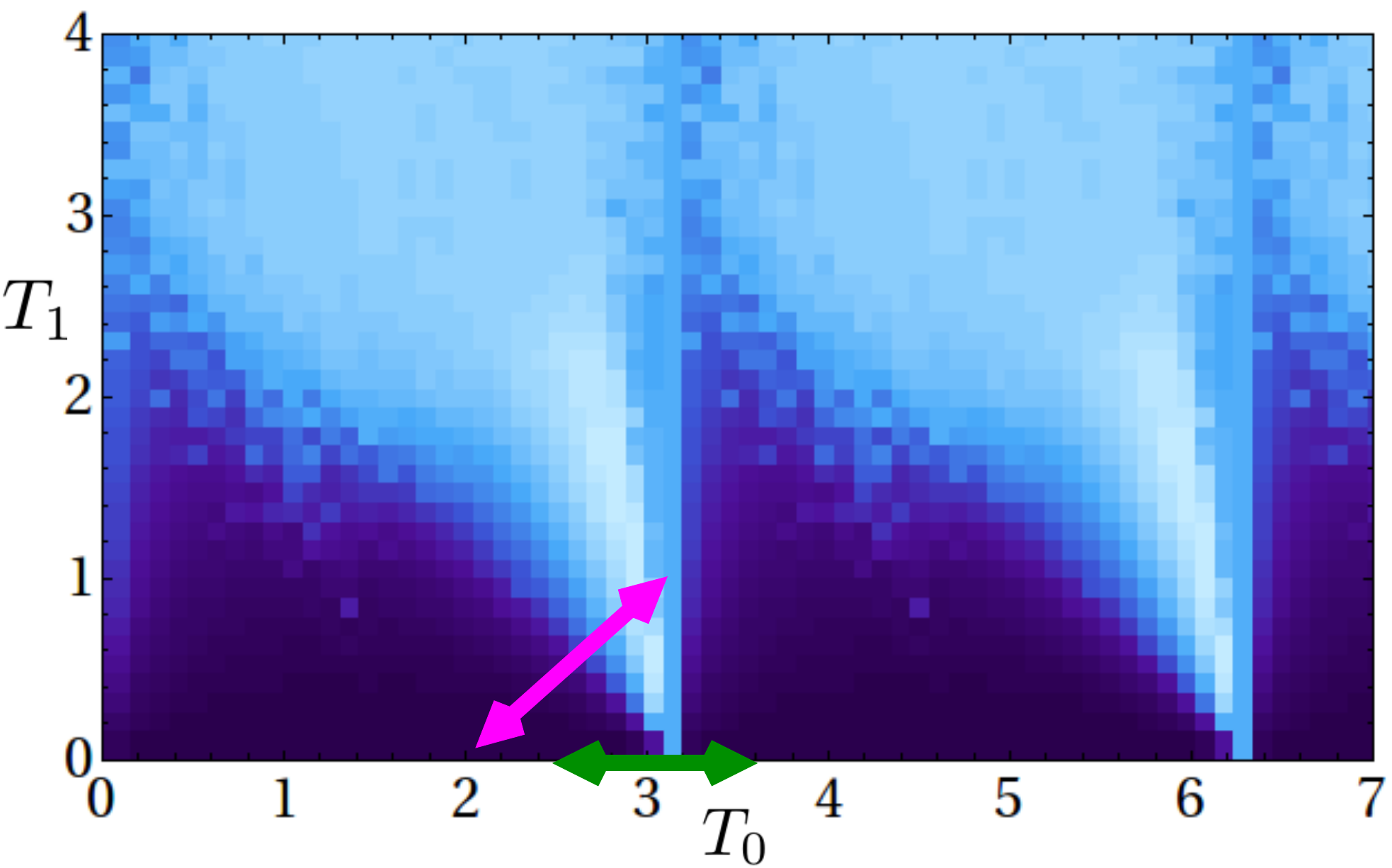}
\caption{\label{fig:T0T1plane} (Color online) Excess energy of the quantum spin chain in the long time limit: $Q=\langle\psi(t)|H_{0}|\psi(t)\rangle_{t\to\infty}-E_{gs}$, where $E_{gs}$ is the ground state energy of the Hamiltonian $H_0$, as a function of the pulse times $T_{0}$ and $T_{1}$ in units of $\hbar/B_x$. Dark blue (light blue) regions correspond to small (large) excess energy. The data is obtained by the exact diagonalization of a spin-$\frac{1}{2}$ chain with $N=15$ spins. }
\end{figure}

In Fig.~\ref{fig:T0T1plane} we show the excess energy in the infinite time as a function of the pulse times $T_{0}$ and $T_{1}$. 
 The dark blue (light blue) regions in parameter space correspond to the system being unable (able) to absorb energy from the time-dependent driving.
The periodicity in $T_{0}$, which is clearly visible in Fig. \ref{fig:T0T1plane}, is due to the periodicity of the spin precession generated by the Hamiltonian $H_{0}$. For
this reason we focus only on $T_{0}\le\pi$. This plot illustrates a sharp crossover between localized and delocalized phases of the driven spin chain as a function of the pulse times. To establish that this crossover becomes a true phase transition in the thermodynamic limit we next show detailed analysis of the excess energy along two line-cuts in the $T_{0}-T_{1}$ plane. Close to the $T_{0}-axis$ (green arrow in Fig.~\ref{fig:T0T1plane}), the effective Hamiltonian $H_{eff}$ can be computed to the first order in $T_{1}$ and all orders in $T_0$ using the Hausdorff-Baker-Campbell formula~\cite{BCH}. For the present problem it is possible to obtain the result of this resummation in the closed analytic form (see Appendix \ref{appendix-resummation}):
\begin{equation}
H_{eff}=H_{av}-\frac{T_{1}}{2T}\left(1-\lambda\cot\left(\frac{\lambda}{2}\right)+\lambda\cot(\lambda)\right)M+\mathcal{O}(T_{1}^{2})
\label{eq:resum}
\end{equation}
where $H_{av}\equiv\frac{1}{T}\left(H_{0}T_{0}+H_{1}T_{1}\right)$ is the time-averaged Hamiltonian, $M$ is a local operator that couples nearest-neighbor and next-nearest-neighbor spins (see Appendix~\ref{appendix-resummation}) and $\lambda=\frac{B_{x}T_{0}}{\hbar}$.

From Eq.~(\ref{eq:resum}) we see that the effective Hamiltonian becomes singular for $T_0=n\,\pi$ (in units of $\hbar/B_x$) no matter how small $T_1$ or the coupling constants in $H_1$ (the $J$s) are. The location of the singularity is also manifestly independent of the system size. As it is seen from the numerical simulations (Fig.~\ref{fig:T0T1plane}), the singularity in the effective Hamiltonian is also manifested in the excess energy of the system in the infinite time limit. We point out that, in the limit of small $T_1$, our system directly extends the kicked rotor model to the many-spin domain. Indeed most of the time the spins precess around the magnetic field $B_x$ getting periodically short kicks by the interacting Hamiltonian $H_1$. Thus in this limit the many-body localization transition directly generalizes the well known kicked rotor localization transition~\cite{chirikov_81, fishman_82}.

Away from the $T_{0}-axis$, nested commutators of order $T_{1}^{n}$ for $n>1$ need to be included in the effective Hamiltonian. These
commutators become difficult to compute analytically at high $n$. Like in the high temperature expansion in statistical physics they involve multiple spin interactions and become more and more delocalized in space. Therefore we have to rely on numerics.
\begin{figure}[ht]
\includegraphics[width=0.9\columnwidth]{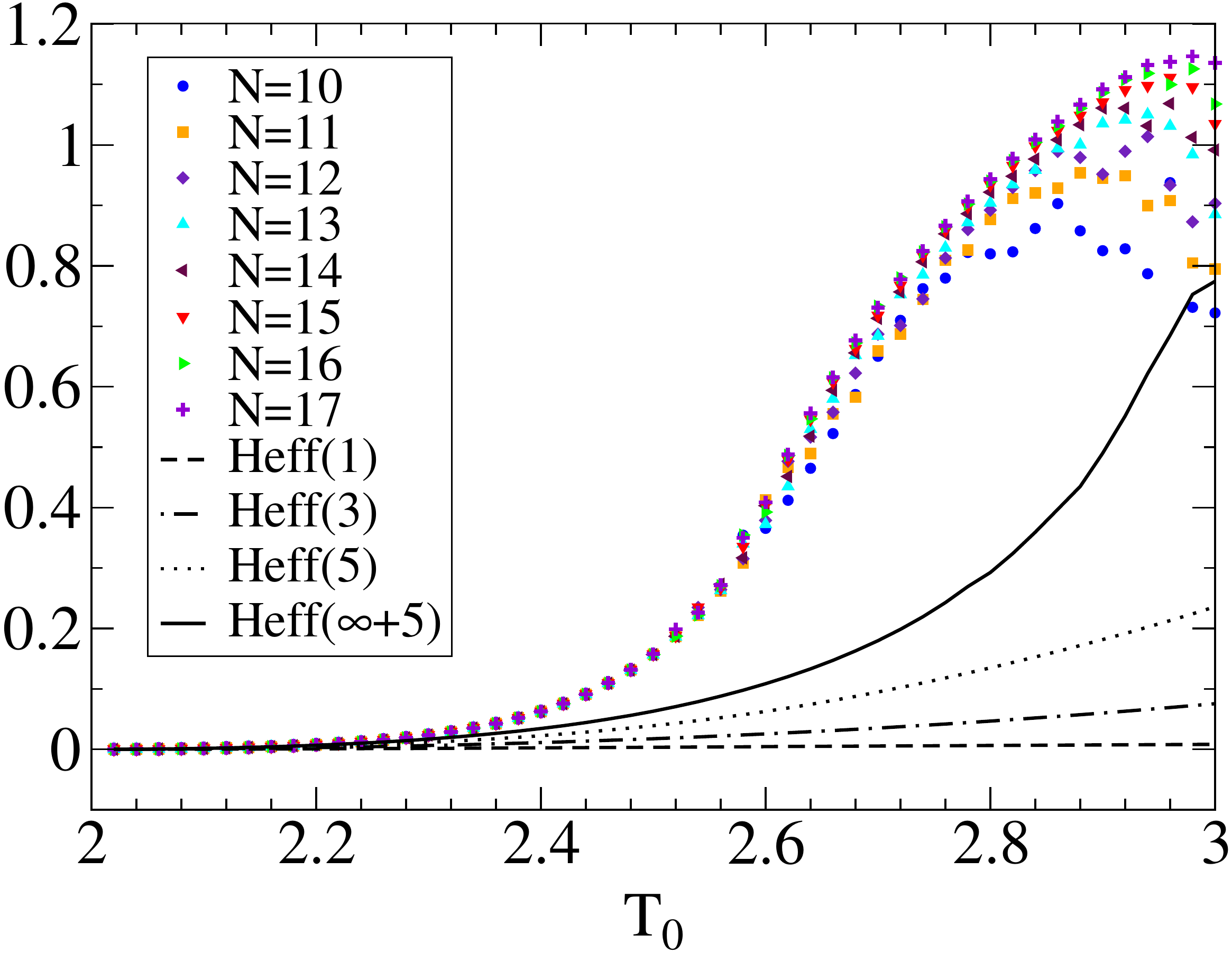}
\caption{(Color online) Asymptotic value of the normalized excess energy, $2\frac{\langle H_{0}\rangle(t=\infty)-E_{gs}}{(E_{max}-E_{gs})}$
where $E_{gs}$ and $E_{max}$ are the lowest and highest eigenvalues of $H_{0}$. The exact results for different system sizes are compared with the predictions obtained by truncating the ME to different orders (see Appendix \ref{appendix-terms-magnus}): $H_{eff}(k)$ contains terms of order $T_{0}^{m}T_{1}^{n}$ with $m+n\le k$ and $H_{eff}(\infty+5)$ denotes the non-perturbative result Eq.~\eqref{eq:resum} together with all the terms $T_{0}^{m}T_{1}^{n}$ with $m+n\le5$. The results from the ME for different system sizes are identical within the image resolution. The date are obtained by exact diagonalization of a quantum spin$-\frac{1}{2}$ chain.
\label{fig:phase-transition}}
\end{figure}
In Fig. \ref{fig:phase-transition} we analyze the long time limit of the excess energy along 
the generic direction $T_{1}=T_{0}-2$ for $2\le T_{0}\le3$ (pink arrow in Fig.~\ref{fig:T0T1plane}).
 
\begin{figure}
\includegraphics[width=0.49\columnwidth]{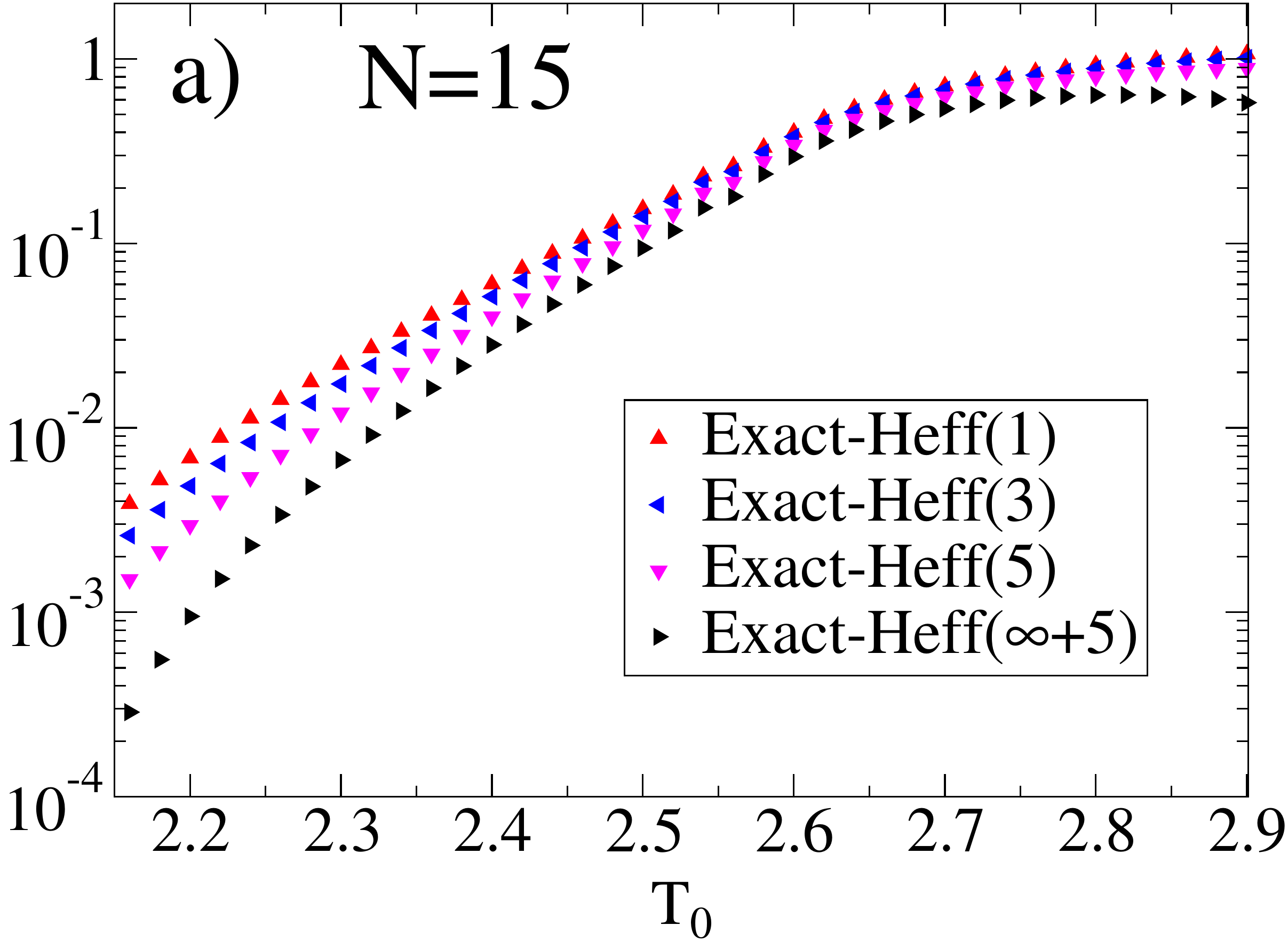}\,\,\,\,\includegraphics[width=0.49\columnwidth]{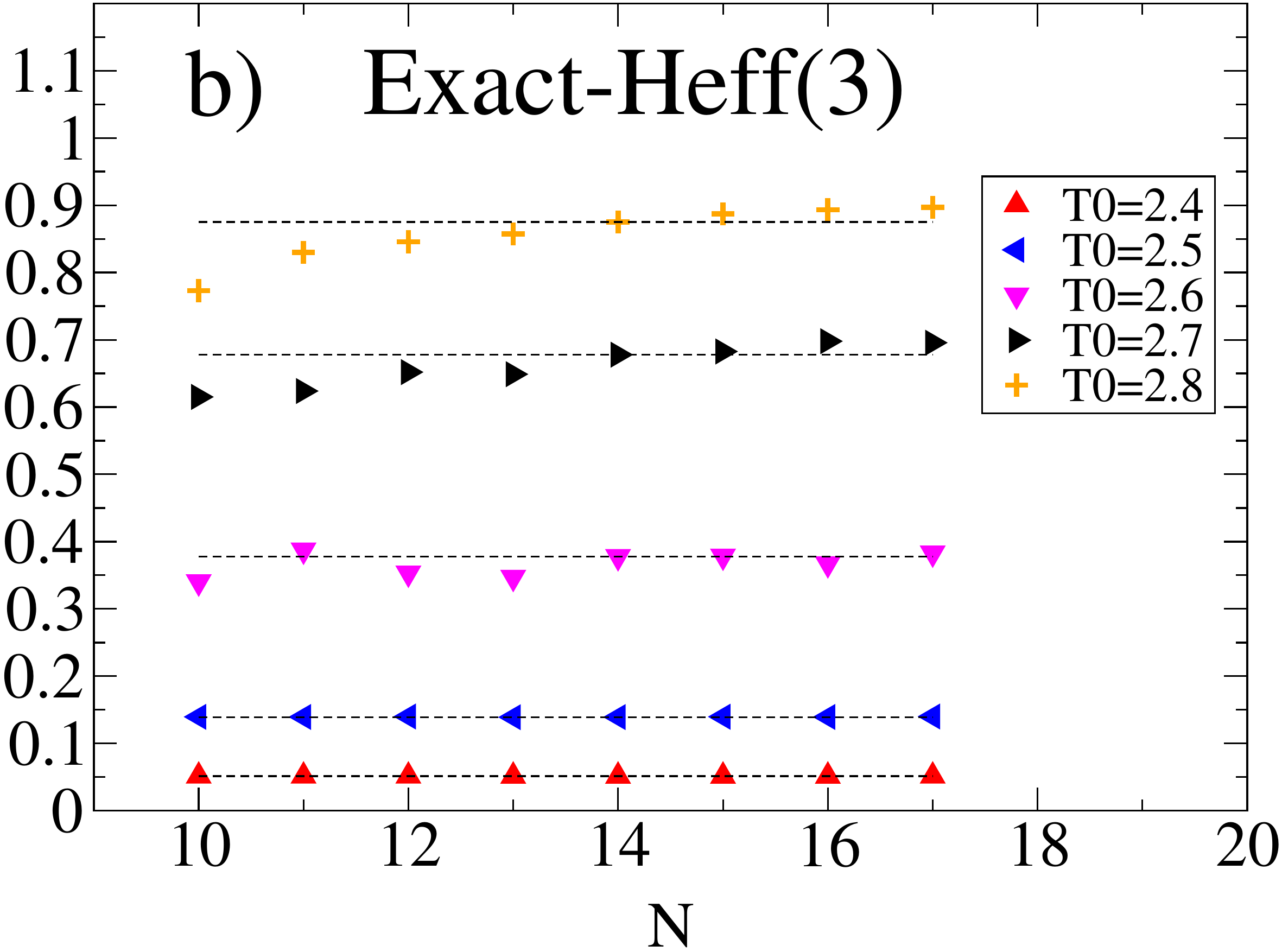}
\caption{(Color online) Difference between the asymptotic value of the exact normalized excess energy and the approximate values obtained by truncating the ME to different orders (see Appendix \ref{appendix-terms-magnus}), for a fixed system size $N=15$ (left panel). Same error for fixed order of truncation versus system size (right panel). The horizontal dotted lines are a guide for an eye. The date are obtained by exact diagonalization of a quantum spin$-\frac{1}{2}$ chain.
\label{fig:error}}
\end{figure}
In Fig. \ref{fig:error} we show the error between the exact asymptotic value of the normalized excess energy (see Fig.~\ref{fig:phase-transition}) and the corresponding excess energies obtained by truncating the ME. From the left panel we see that, as expected, for short periods the ME becomes asymptotically exact. From the right panel we see that the truncation error at short periods does not increase with the system size indicating existence of the localized phase in the thermodynamic limit. The latter implies existence of the {\it localization-delocalization transition}. For long periods (see Fig.~\ref{fig:phase-transition}) the energy clearly approaches the infinite temperature asymptotic value.
From these data we can estimate that the value of the critical period (where the ME breaks down) is $T_c \approx 2.6$.

\subsection{Diagonal Entropy of the quantum spin chain \label{sec-diag-en}}

\begin{figure}[ht]
\includegraphics[width=0.49\columnwidth]{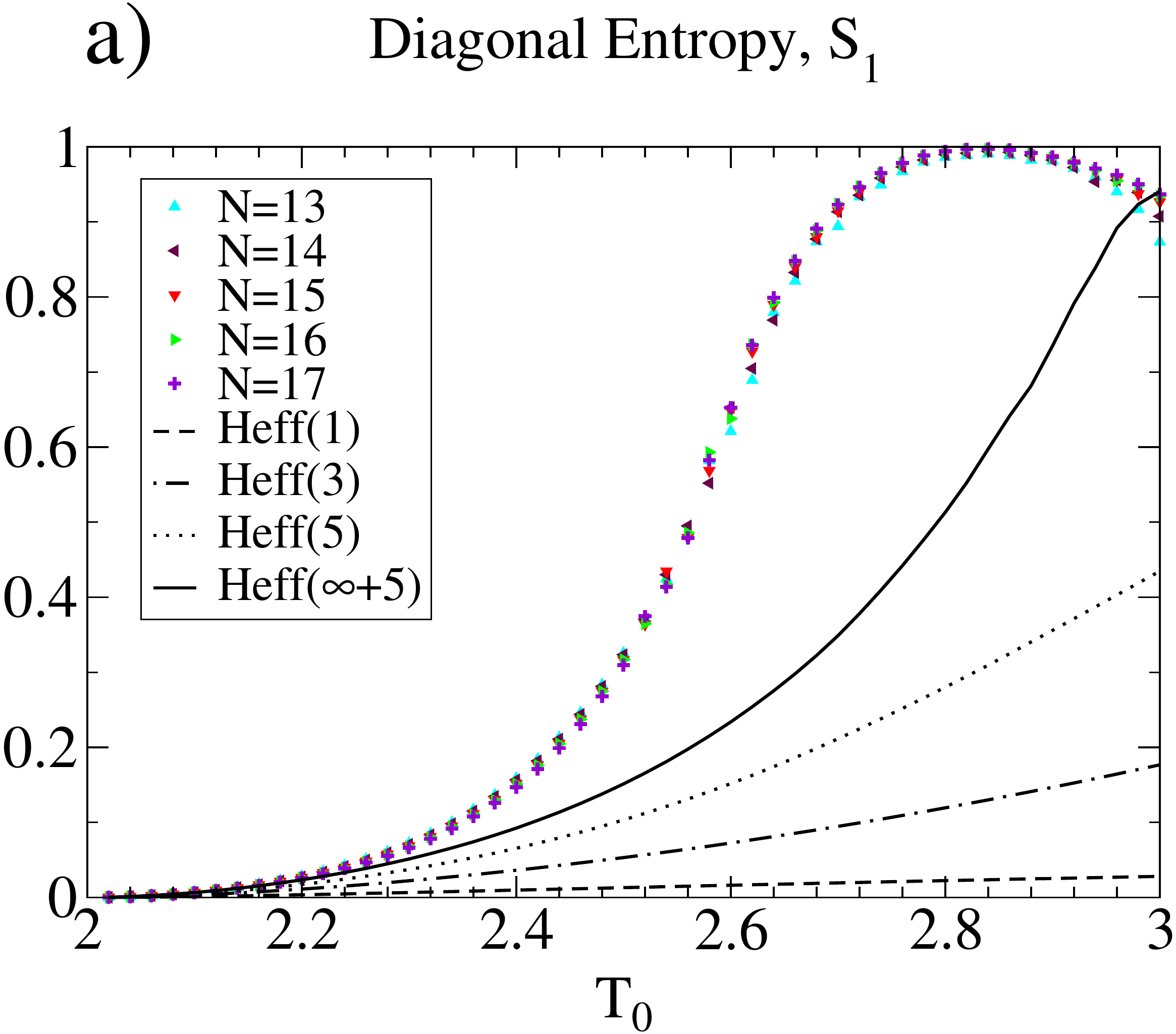}\,\,\,\,\includegraphics[width=0.49\columnwidth]{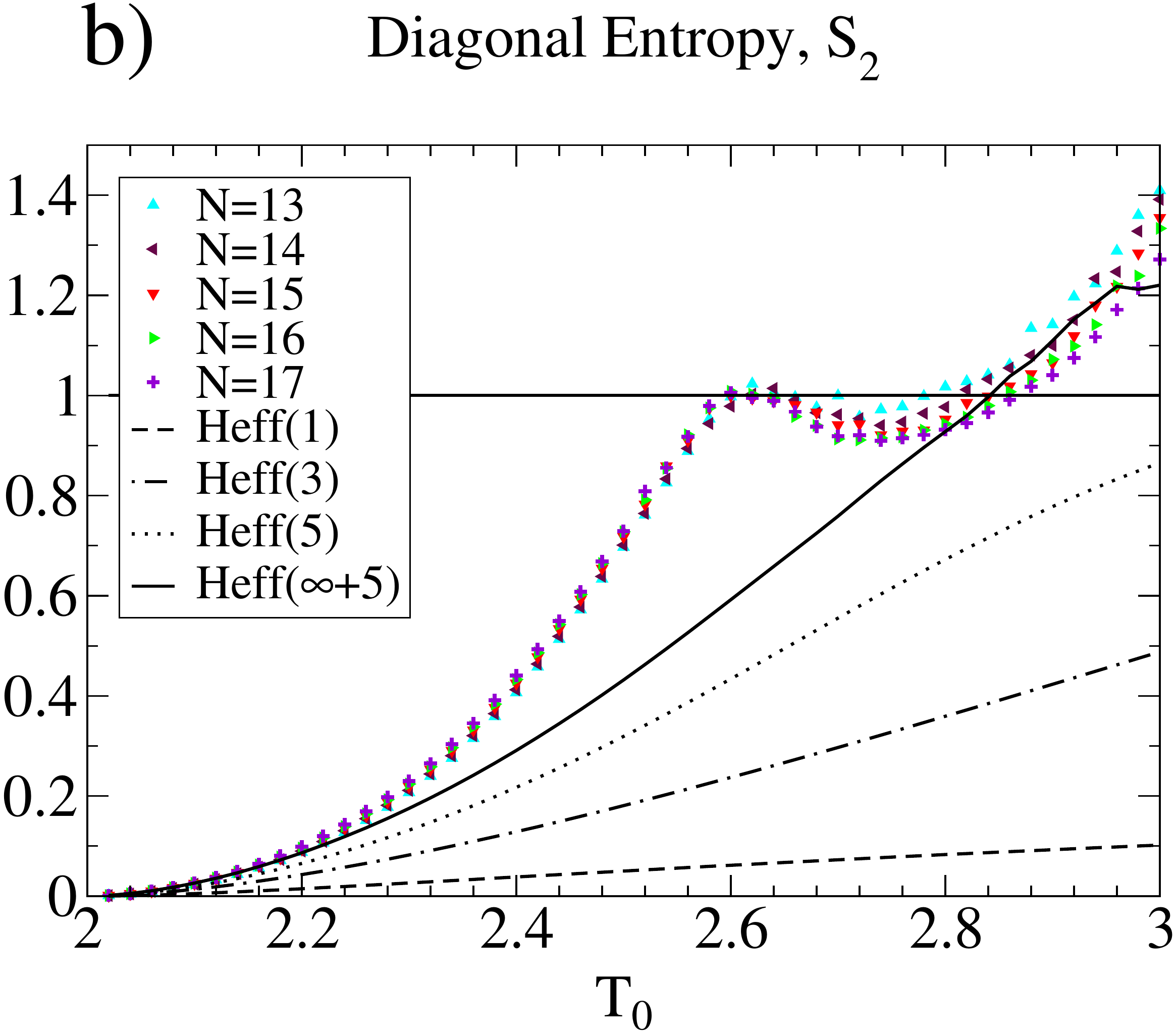}
\caption{(Color online) Asymptotic value of the normalized diagonal entropies $S_1$ and $S_2$ (see Eq.~\eqref{s1-s2}). The exact results for different system sizes are compared with the predictions obtained by truncating the ME to different orders  (see Appendix \ref{appendix-terms-magnus}): $H_{eff}(k)$ contains terms of order $T_{0}^{m}T_{1}^{n}$ with $m+n\le k$ and $H_{eff}(\infty+5)$ denotes the non-perturbative result in $T_0$, Eq.~\eqref{eq:resum}, together with all the other terms $T_{0}^{m}T_{1}^{n}$ with $m+n\le5$. The results from the ME for different system sizes are identical within the image resolution. In the right panel, the horizontal line at value $1$ is a guide for the eye. The data are obtained by exact diagonalization of a quantum spin$-\frac{1}{2}$ chain. 
\label{entropy-pt}}
\end{figure}

The diagonal entropy ~\cite{ap_ent} serves as a measure of the occupation in the Hilbert space. 
It is defined as $S(t)=-\sum_{k}p_{0}^{k}(t)\log p_{0}^{k}(t)$ where $p_{0}^{k}(t)$ are the occupation probabilities on a complete orthonormal basis of the Hilbert space. The value of the diagonal entropy is basis-dependent. To study the energy-delocalization transition it is convenient to choose the basis of the eigenstates of $H_0$. However the spectrum of $H_0$ is highly degenerate and the probabilities $p_{0}^{k}(t)$ depend on the (arbitrary) choice of basis in each degenerate subspace of $H_0$\footnote{We thank the anonymous referee for drawing our attention to this issue}. 
On the other hand, the occupation probabilities of the each degenerate subspace, i.e. the probability of an outcome of a measurement of $H_0$, are independent of this choice and are given by $p(E_0)(t)=\sum_{k=1}^{N(E_0)} p_{0}^{k}(t)$ 
where $E_0$ are the distinct eigenvalues of $H_0$ and the sum is over the states in each degenerate subspace of dimension $N(E_0)$. 

We define two diagonal entropies, $S_1$ and $S_2$, which differ solely by the choice of basis in each degenerate subspace:
\begin{equation}
S_1=-\frac{1}{\mathcal{N}_1}\sum_{E_0}p(E_0)\ln\left(\frac{p(E_0)}{N(E_0)}\right),\quad S_2=-\frac{1}{\mathcal{N}_2}\sum_{E_0}p(E_0)\ln p(E_0),
\label{s1-s2}
\end{equation}
The first entropy, $S_1$, corresponds to the situation in which each state in the degenerate subspace has the same occupation ($p_{0}^{k_j}=p(E_0)/N(E_0)$ for $j=1,...,N(E_0)$). The second entropy, $S_2$, corresponds to the situation in which a single state in the degenerate subspace carries the entire probability ($p_{0}^{k_1}=p(E_0)$ and $p_{0}^{k_j}=0$ for $j=2,..,N(E_0)$). Both entropies correspond to the the general definition of diagonal entropy for two different choices of the basis in the degenerate subspaces. 
The two entropies provide complementary information since $S_1$ weighs all the states in the Hilbert space equally while $S_2$ weighs all subspaces equally irrespectively of their size. Note that $S_2$ is equal to the sum of entanglement entropies of each degenerate subspace. The normalization factors, $\mathcal{N}_1$ and $\mathcal{N}_2$, are chosen so that an infinite temperature state, i.e. a random superposition of all states in the Hilbert space, has entropy $1$. They are given by:
\begin{equation*}
\mathcal{N}_1=\ln N_{tot},\quad \mathcal{N}_2=-\sum_{E_0}\tilde{p}(E_0)\ln \tilde{p}(E_0)
\end{equation*}
where $N_{tot}\equiv\sum_{E_0} N(E_0)$ is the total number of states in the Hilbert space and $\tilde{p}(E_0)\equiv N(E_0)/N_{tot}$ is the expected occupation probability (for an infinite temperature state) of each degenerate subspace. 
Note that in the thermodynamic limit $\mathcal{N}_1$ is extensive ($\mathcal{N}_1\approx L$) while $\mathcal{N}_2$ increases logarithmically with the system size ($\mathcal{N}_2 \approx \ln(L)/2$).

In Fig. \ref{entropy-pt} we study the behavior of these two entropies along the generic line $T_1 = T_0 - 2$ for $2 \le T_0 \le3$. In particular we show the asymptotic value of the diagonal entropy for different system sizes and we compare it to the values obtained by truncating the ME to different orders. The asymptotic values have been computed by projecting the initial state to the eigenstates of $H_{eff}$  and then back to $H_0$. This procedure is equivalent to the assumption of infinite time averaging with respect to $H_{eff}$ and the asymptotic values obtained correspond to the prediction of the diagonal ensemble of $H_{eff}$. The left panel of Fig.~\ref{entropy-pt} shows that $S_1$ behaves similarly to the excess energy (see Fig.~\ref{fig:phase-transition}). In particular, we see that for short (long) periods the prediction of the ME converges (does not converge) to the exact result obtain by full-diagonalization. We also note that for short (long) periods the normalized entropy decreases (increases) with the system size. This suggests that we can locate the transition by finding the crossing of $S_1$ for different system sizes. In agreement with the data reported in Fig.~\ref{fig:error}, the transition seems to be located at $T_c\approx 2.6$. In the right panel of Fig.~\ref{entropy-pt} we show the entropy $S_2$. We see again that for short (long) periods the ME converges (does not converge) to the exact results. Moreover the entropy $S_2$ has an interesting non-monotonic behavior and displays a local maximum at $T\approx 2.6$ where $S_2\approx1$ which is the value expected for the fully delocalized (i.e. infinite temperature) state. This fact again suggest that the critical value of the period is $T_c\approx 2.6$. The behavior of $S_2$ for $T>T_c$ is not fully understood and might be dominated by finite size effects. We note that for the largest system size available $S_2$ tends to flatten and therefore we can conjecture that in the thermodynamic limit $S_2$ approaches the constant value $1$ for all $T>T_c$. However the system sizes available are too small to support this conjecture. 
We also note that by breaking the integrability of $H_0$ the degeneracies are lifted and the diagonal entropy becomes uniquely defined. 
We expect this latter diagonal entropy to qualitatively behave as $S_1$. In conclusion, the numerical evidences reported in Fig.~\ref{entropy-pt} show that the energy increase observed in Fig.~\ref{fig:phase-transition} is indeed caused by a delocalization of the system in Hilbert space.

\subsection{Time series of the Energy and Entropy of the quantum spin chain}
\begin{figure}[ht]
\includegraphics[width=0.49\columnwidth]{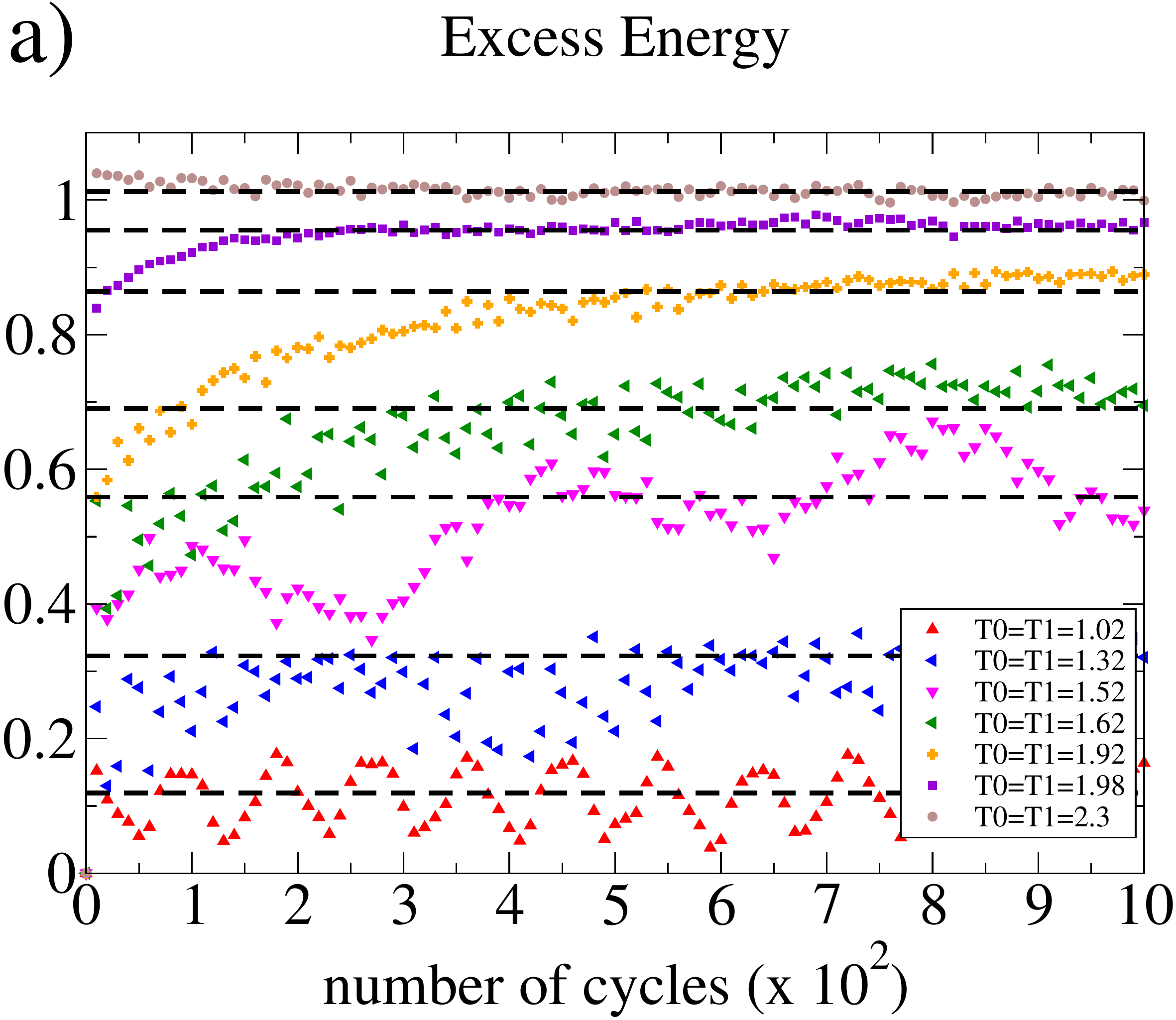}\,\,\,\,\includegraphics[width=0.49\columnwidth]{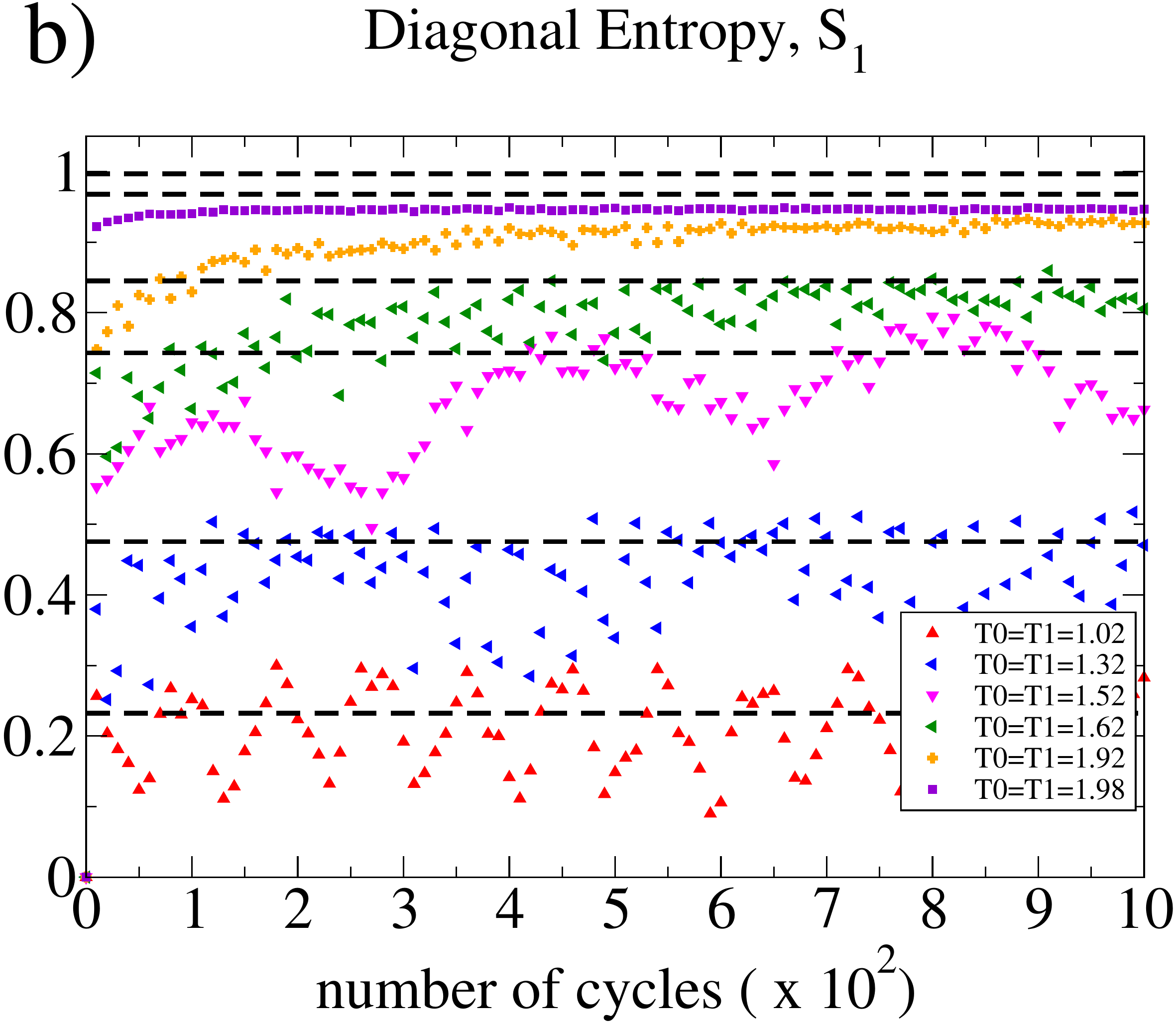}
\caption{(Color online) Time evolution of (left) the excess energy for spin $Q=2\frac{\langle H_{0}\rangle(t=\infty)-E_{gs}}{(E_{max}-E_{gs})}$
where $E_{gs}$ and $E_{max}$ are the lowest and highest eigenvalues of $H_{0}$ 
and (right) the normalized diagonal entropy, $S_1(t)$ (see Eq.~\eqref{s1-s2}). The dotted lines indicate the predictions from the diagonal ensemble. The date are obtained by exact diagonalization of a quantum spin$-\frac{1}{2}$ chain with $N=17$ spins.  
\label{S2}}
\end{figure}

In this section we report our numerical findings for the time-evolution of the excess energy and the diagonal entropy $S_1$ (see Eq.~\eqref{s1-s2}) of the quantum chain with $N=17$ spins for different values of the pulse times $T_0=T_1=T/2$. In particular, in Fig.~\ref{S2} we compare the time series of these quantities with the predictions from the diagonal ensemble of $H_{eff}$. 

On the left panel we observe that the excess energy, $Q=2\frac{\langle H_{0}\rangle(t=\infty)-E_{gs}}{(E_{max}-E_{gs})}$, approaches the diagonal ensemble predictions for all the values of the pulsed times $T_0$ and $T_1$. 
Moreover, we see that for short and long periods (such as $T_0=T_1=1.02$ and $T_0=T_1=2.3$ respectively) the asymptotic value of the excess energy is approached quickly while for intermediate periods (such as $T_0=T_1=1.92$) this approach is extremely slow. This seems to suggest that the localization transition is associated with a divergent time scale, i.e. the time to approach the asymptotic value.
On the right panel we observe a similar behavior for the diagonal entropy $S_1$ (see Eq.~\eqref{s1-s2}). In this case however the prediction from the diagonal ensemble seems to over-estimate, for long periods, the value of the diagonal entropy. This small discrepancy, which is  non-extensive and thus vanishes in thermodynamic limit, is caused by the fact that the entropy is not a linear function of the density matrix and can be fully explained by a more careful analysis \cite{ikeda}. Finally we note that for long periods the fluctuations in the time-series of the both observables (excess energy and diagonal entropy) are suppressed. This is expected since in the delocalized phase more states participate in the dynamics and the statistical average is improved.

\subsection{Classical spin chain}
\begin{figure}
\includegraphics[width=0.49\columnwidth]{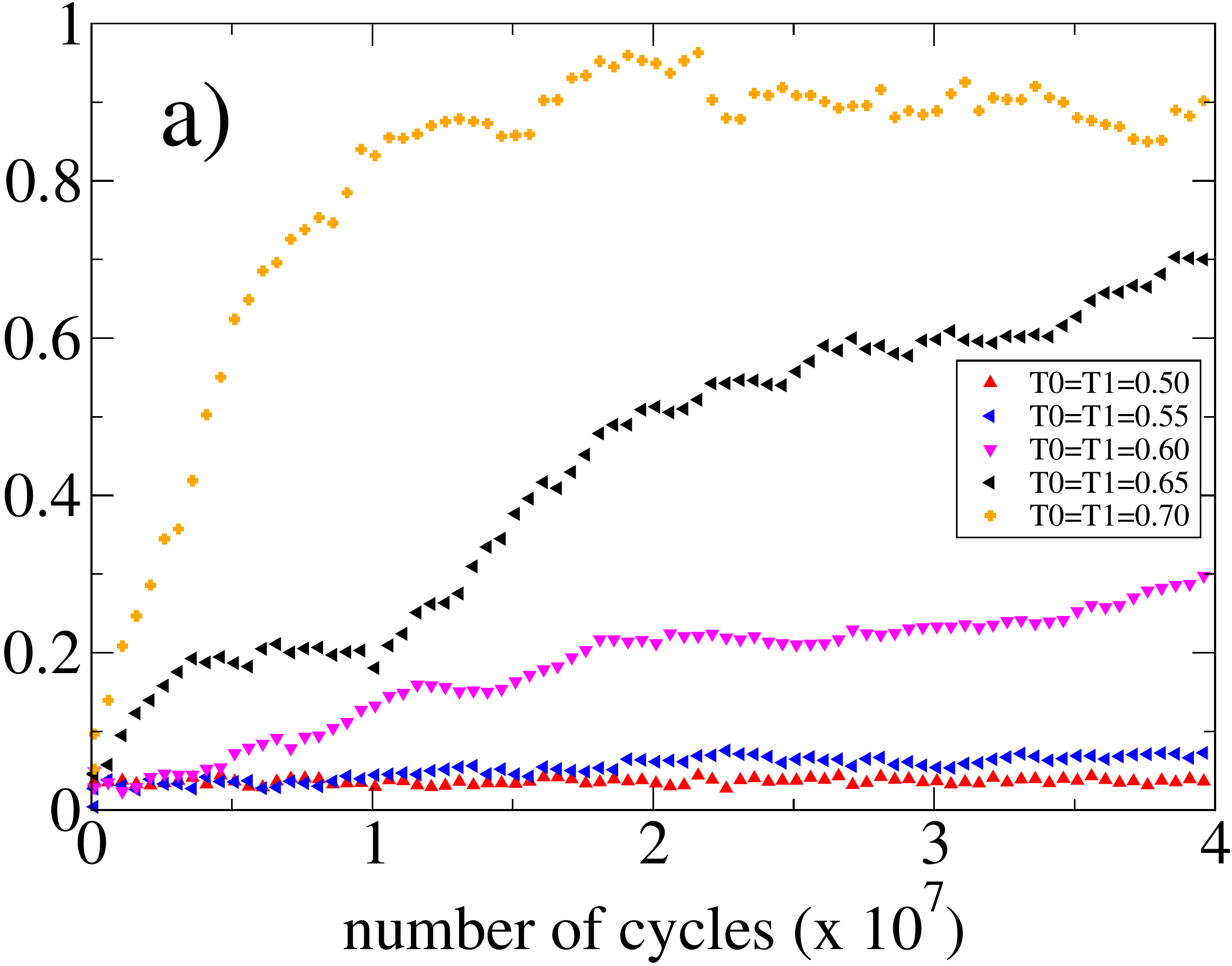}\,\,\,\,\includegraphics[width=0.49\columnwidth]{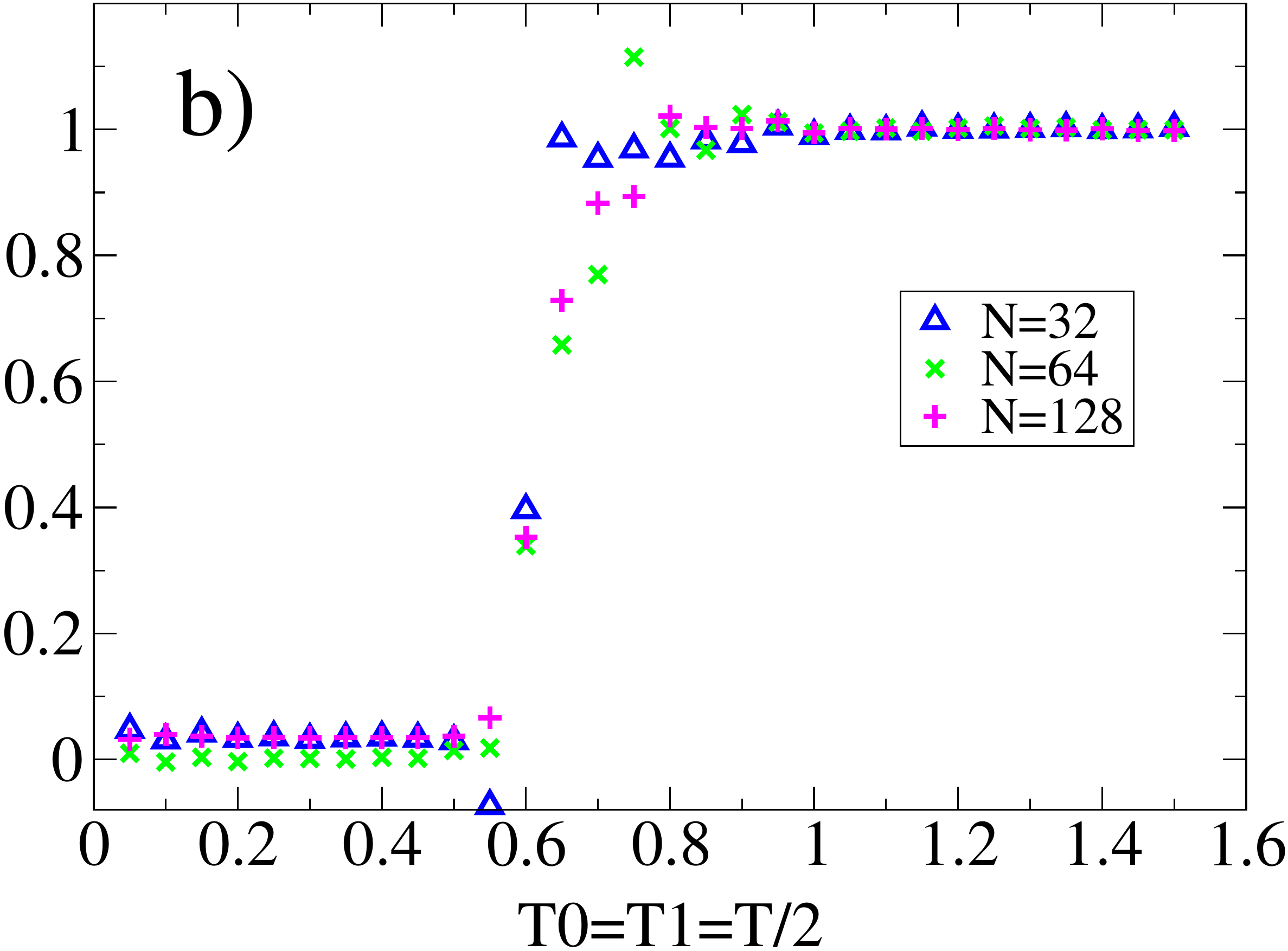}
\caption{(Color online) Time evolution of the excess energy for spin, $Q\equiv\frac{E(t)-E_{gs}}{N/2}$, for a classical chain with $N=128$ spins (left panel) and asymptotic excess energy for different systems sizes (right panel).
\label{fig:classical}}
\end{figure}

The localization transition is also found for the system of classical spins (see Fig.~\ref{fig:classical}). On the left panel we show the time evolution of the excess energy for different values of the pulse times $T_0$ and $T_1$. We observe that close to the transition the evolution of the excess energy is extremely slow (note the scale in Fig.~\ref{fig:classical}) suggesting that the this transition is characterized by a divergent time scale, i.e. the time required to reach the asymptotic state. On the right panel we show the asymptotic value of the excess energy for different system sizes. These asymptotic values have been computed by averaging the energy over the last $10^7$ cycles of the evolution. 
However close to the transition, it is not clear if the energy has reached its asymptotic value. So more work is needed to check whether the sharp crossover in classical systems seen on the right panel of Fig.~\ref{fig:classical} becomes a true phase transition in the long time limit. The analytical argument based 
on the resummation of the Hausdorff-Baker-Campbell formula (Eq.~\eqref{eq:resum}) suggests this is in fact a transition.

\section{Conclusions}

We presented analytic and numerical arguments suggesting the existence of the localization transition in interacting driven systems in the thermodynamic limit. For short periods of driving $T<T_{c}$ the system is unable to absorb energy beyond some threshold even in the infinite time limit while for longer periods $T>T_{c}$ the system absorbs energy and  becomes delocalized in the entire Hilbert space. This interpretation is confirmed by the study of the diagonal entropy~\cite{ap_ent} (see Sec. \ref{sec-diag-en}). The diagonal entropy serves as a measure of the occupation of the Hilbert space. For short periods it remains bounded at all times and for long periods, up to small sub-extensive corrections, approaches the value expected for a completely delocalized (i.e infinite temperature) state. We associate the delocalization transition with the divergence of the short time ME for the effective Floquet Hamiltonian similarly to the divergence of the high-temperature expansion in statistical physics. In this sense this transition is reminiscent of the divergence of the short time expansion in the Loschmidt echo recently obtained for a quench in the transverse field Ising spin chain~\cite{heyl_12}. 
The precise characterization of the two phases and of the nature of the transition and its connection with the radius of convergence of the ME will be investigated in a future publication. While at the moment we can not rigorously prove that this is a true phase transition, the right panel in Fig.~\ref{fig:error} gives very convincing evidence that at short periods the energy always remains bounded even in the thermodynamic limit indicating the existence of the localized phase. We believe that this transition is generic for systems with bounded single particle excitations. In fact, when the single particle spectrum is unbounded the ME does not converge in the limit $T\rightarrow 0$ \cite{Kehrein_1} and the existence of the energy-localized phase is unlikely. The physical reason is that, for an unbounded spectrum, the periodic driving can generate excitation of arbitrarily high energy, $E_{ex}\sim 2\pi/T$, (corresponding to the absorption of quanta of the driving frequency) which are not described by the effective time-independent Hamiltonian. However it can be shown that, for systems with a finite single particle bandwidth, the ME converges in the $T\rightarrow 0$ limit \cite{Kehrein_1,Kehrein_2}. For these systems, we expect to observe the energy-localization transition as a function of the driving period.

This transition should be experimentally detectable in isolated interacting systems like cold atoms, cold ions, nuclear spins and various materials with interacting spin degrees of freedom.

\section{Acknowledgements}
We are grateful to Kai Ji and Boris V. Fine for interesting conversations in the early stage of this project and for sharing their preliminary results with us. We also thank G. S. Boutis, M. Heyl, S. Kehrein, P. Krapivsky, P. Metha, V. Oganesyan and M. Rigol for interesting conversations. This work was partially supported by BSF 2010318, NSF DMR-0907039, NSF PHY11-25915, AFOSR FA9550-10-1-0110, as well as the Simons and the Sloan Foundations. The authors are thankful for the hospitality of the Kavli Institute for Theoretical Physics at UCSB, where part of the work was done.

\appendix 
\renewcommand*{\thesection}{\Alph{section}}

\section{Appendix: Magnus Expansion for the Kapitza Pendulum \label{appendix-kapitza}}
Here we will show how to apply the Magnus Expansion (ME) to the Kapitza pendulum problem.
In particular we will show how to derive the dynamical stabilization condition, first obtained by Kapitza.
The quantum time-dependent Hamiltonian for the Kapitza pendulum is: 
\begin{equation}
\hat{H}(t)=\frac{1}{2m}\hat{p}_{\theta}^{2}+f(t)\cos\hat{\theta}
\end{equation}
where $f(t)=-m\left(\omega_{0}^{2}+\frac{a}{l}\gamma^{2}\cos\left(\gamma t\right)\right)$ and $\hat{\theta},\,\,\hat{p}_{\theta}$
are quantum operators with canonical commutation relations $\left[\hat{\theta};\hat{p}_{\theta}\right]=i\hbar$.
The explicit form of the first three terms in the ME are (see the review article~\cite{Magnus-report}):
\begin{equation}
\begin{array}{c}
\hat{H}_{eff}^{(1)}=\frac{1}{T}\int\hat{H}(t_{1})\\
\hat{H}_{eff}^{(2)}=\frac{1}{2T\left(i\hbar\right)}\iint\left[\hat{H}(t_{1});\hat{H}(t_{2})\right]\\
\hat{H}_{eff}^{(3)}=\frac{1}{6T\left(i\hbar\right)^{2}}\iiint\left(\left[\hat{H}(t_{1});\left[\hat{H}(t_{2});\hat{H}(t_{3})\right]\right]+\left[\hat{H}(t_{3});\left[\hat{H}(t_{2});\hat{H}(t_{1})\right]\right]\right)
\end{array}
\end{equation}
where the time integration domains are ordered, i.e. $0<t_{n}<t_{n-1}<...<t_{1}<T$.
Recalling that the period of the driving is $T=\frac{2\pi}{\gamma}$ after some simple algebra we obtain:
\begin{equation}
\begin{array}{c}
\hat{H}_{eff}^{(1)}=\frac{1}{2m}\hat{p}_{\theta}^{2}-m\omega_{0}^{2}\cos\hat{\theta}\\
\hat{H}_{eff}^{(2)}=0\\
\hat{H}_{eff}^{(3)}=-\left(\frac{1}{4m}\,\frac{a}{l}\right)\frac{\left[\hat{p}_{\theta}^{2};\left[\hat{p}_{\theta}^{2};\cos\hat{\theta}\right]\right]}{\left(i\hbar\right)^{2}}+ \frac{m}{2} \left( \frac{a}{l} \omega_{0}^{2}-\left(\frac{\gamma a}{2 l}\right)^2\right)\frac{\left[\cos\hat{\theta};\left[\hat{p}_{\theta}^{2};\cos\hat{\theta}\right]\right]}{\left(i\hbar\right)^{2}}
\end{array}
\label{eq:ME-quantum}
\end{equation}
Substituting the explicit value for the commutators in $\hat{H}_{eff}^{(3)}$ we obtain:
\begin{equation}
\hat{H}_{eff}^{(3)}=\left(\frac{1}{4m}\,\frac{a}{l}\right)\left(\hat{p}_{\theta}^{2}\cos\hat{\theta}+2\hat{p}_{\theta}\cos\hat{\theta}\hat{p}_{\theta}+\cos\hat{\theta}\hat{p}_{\theta}^{2}\right)+m\left(\left(\frac{a \gamma}{2 l}\right)^2-\frac{a}{l}\omega_{0}^{2} \right)\sin^{2}\hat{\theta}
\label{eq:Kapitza-quantum}
\end{equation}
Combining Eqs.~\eqref{eq:ME-quantum} and Eq.~\eqref{eq:Kapitza-quantum} we obtain the first three terms in ME for the quantum Kapitza pendulum.
Up to this order,  the classical ME (see Eq.~(6) in the main text) can be obtained from the quantum counterpart by substituting the quantum operators with classical variables. This is not true in general and a more rigorous approach is necessary to derive the classical limit of the ME. 

Before showing the general approach to obtain the classical limit of the ME we note that the first three terms in the ME suffice to explain the dynamical stabilization of the classical Kapitza pendulum. Let us assume that $\frac{a}{l}\omega_{0}^{2}\ll\left(\frac{a \gamma}{2 l}\right)^2$ (we will check this assumption a posteriori) then by collecting the terms in $H_{eff}^{(1)}$ and $H_{eff}^{(3)}$ that involve only the coordinates we obtain the effective potential $U_{eff}=m\omega_{0}^{2}\left(-\cos\theta+\left(\frac{a\gamma}{2l\omega_{0}}\right)^{2}\sin^{2}\theta\right)$ which develops a new minimum at the inverted position, $\theta=\pi$, 
for $\frac{a}{l}\,\frac{\gamma}{\omega_{0}}>\sqrt{2}$. Moreover close to the stabilization 
transition $\left(\frac{a\gamma}{l}\right)^{2}\sim2\omega_{0}^{2}$ and we can check that our assumption is justified as long as 
$\frac{a}{l}\ll\frac{1}{2}$, which is one of the condition required by the original derivation by Kapitza.

Now let us discuss how one can obtain obtain the classical limit of the ME through the phase-representation of Quantum Mechanics.  Let us briefly review the formalism~\cite{ap_phasespace}. 
In the phase-space representation of quantum mechanics the quantum operator $\hat{\Omega}$ is replaced by its Weyl Symbol which is a classical function over the phase-space variables $x$ and $p$ (which can be vectors for multi-dimensional problems). 
The Weyl symbol is defined as 
\begin{equation}
\Omega_{w}(x,p)=\int ds\,\langle x-\frac{s}{2}|\hat{\Omega}|x+\frac{s}{2}\rangle\exp\left[\frac{i}{\hbar}p\cdot s\right]
\end{equation}
If the quantum operator $\hat{\Omega}(\hat{x},\hat{p})$ is written in the symmetrized form then it can be shown 
that its Weyl symbol $\Omega_{w}$ is simply obtained by the substitution $\hat{x}\rightarrow x$ and $\hat{p}\rightarrow p$.
In particular, this is true for all operators of the form $\hat{\Omega}(\hat{x},\hat{p})=\hat{A}(\hat{x})+\hat{B}(\hat{p})$. 
In this representation the commutator of two quantum operators is written in terms of the Moyal Bracket:
\begin{equation}
\left[\hat{\Omega}_{1};\hat{\Omega}_{2}\right]\rightarrow i\hbar\left\{ \Omega_{1,w};\Omega_{2,w}\right\} _{MB}\equiv i\hbar\,\,\Omega_{1,w}\left(-\frac{2}{\hbar}\sin\left(\frac{\hbar}{2}\Lambda\right)\right)\Omega_{2,w}
\end{equation}
where $\Lambda$ is the differential operators $\ensuremath{\Lambda\equiv\overleftarrow{\partial_{p}}\,\,\overrightarrow{\partial_{x}}-\overleftarrow{\partial_{x}}\,\,\overrightarrow{\partial_{p}}}$. By expanding $-\frac{2}{\hbar}\sin\left(\frac{\hbar}{2}\Lambda\right)$ in powers of $\hbar$ we obtain the classical limit, i.e. the Poisson Bracket (zero order in $\hbar$), and the quantum correction (higher powers of $\hbar$). Note that the first quantum correction is proportional to $\Lambda^{3}$. Thus if we are interested in the classical limit of ME ($\hbar\rightarrow0$) it suffices to truncate the expansion to order $\Lambda$. 
This amounts to the replacement of the quantum commutator with the Poisson Bracket, i.e. $\frac{1}{i\hbar}[...]\rightarrow\{...\}$. 
In the case of Kapitza pendulum, the only non-zero contribution to the first few nested commutators comes from the order $\Lambda$ and the classical and quantum ME are identical. This is not always true since at higher order in the ME terms of order $\Lambda^{3}$ (or higher) need to be considered and nontrivial differences between the classical and quantum expansions will be generated.

\section{Appendix: Resummation of the Hausdorff-Baker-Campbell formula \label{appendix-resummation}}
Here, we show how to re-sum the Hausdorff-Baker-Campbell formula (HBC)~\cite{BCH} to obtain the Hamiltonian~\eqref{eq:resum}.
First, it is convenient to introduce the shorthand notations $X\rightarrow  -\frac{i}{2\hbar}H_0T_0,\,\, Y\rightarrow -\frac{i}{\hbar}H_1T_1$ and $Z\rightarrow -\frac{i}{\hbar}H_{eff}T$ where $T=T_0+T_1$. With these notations the effective Hamiltonian for the protocol considered (see Fig.~\ref{fig:protocol}) is defined by $Z\equiv\log\left[e^{X}e^{Y}e^{X}\right]$. Expanding this expression to the first order in $Y$ we obtain the HBC formula:
\begin{equation}
Z=2X+Y-\sum_{n=1}^{\infty}\frac{2(2^{2n-1}-1)}{(2n)!}\, B_{2n}\,\left\{ X^{2n};Y\right\} +\mathcal{O}(Y^{2})
\label{eq:HBC}
\end{equation}
where $B_{k}$ are the first Bernoulli numbers ($B_{0}=1,B_{2}=\frac{1}{6},B_{4}=-\frac{1}{30}, ...$) 
and $\left\{ X^{k};Y\right\} \equiv\left[X;\left[X;...\left[X;Y\right]\right]\right]$ is the nested commutator 
with $k$ operators $X$ and only one operator $Y$ appearing at the right most position. 
Note that Eq.~\eqref{eq:HBC} is different from the expression usually reported 
in the literature, where the less symmetric situation $Z\equiv\log\left[e^{X}e^{Y}\right]$ is considered.
The simple form of the Hamiltonian $H_0$ allows us to compute all the nested commutators analytically. 
In fact, using the explicit form of $H_0$ and $H_1$ (see Eq.~\eqref{eq:Hamiltonians}) together with the commutation relations for spin 
operators $[s_{n}^{\alpha};s_{m}^{\beta}]=i\,\delta_{n,m}\epsilon_{\alpha\beta\gamma}s_{n}^{\gamma}$ , it is easy to verify that
\begin{equation}
\begin{array}{c}
\left\{ \left(H_{Bx}\right)^{2n};H_{z}\right\} =2^{(2n-1)}W_{zz-yy},\,\,\,
\left\{ \left(H_{Bx}\right)^{2n};H_{z}^{\prime}\right\} =2^{(2n-1)}W_{zz-yy}^{\prime}\\
\left\{ \left(H_{Bx}\right)^{2n};H_{\parallel}\right\} =-2^{(2n-1)}W_{zz-yy},\,\,\,
\left\{ \left(H_{Bx}\right)^{2n};H_{\parallel}^{\prime}\right\} =-2^{(2n-1)}W_{zz-yy}^{\prime}
\end{array}
\end{equation}
where $W_{zz-yy}\equiv\sum_{n}\left(s_{n-1}^{z}s_{n}^{z}-s_{n-1}^{y}s_{n}^{y}\right)$ 
and $W_{zz-yy}^{\prime}\equiv\sum_{n}\left(s_{n-2}^{z}s_{n}^{z}-s_{n-2}^{y}s_{n}^{y}\right)$.
Plugging these expressions in Eq.~\eqref{eq:HBC} and performing the sum we find Eq.~\eqref{eq:resum} 
where $M\equiv \left( J_{z}-J_{\parallel} \right) W_{zz-yy}+ \left( J_{z}^{\prime}-J_{\parallel}^{\prime} \right)W_{zz-yy}^{\prime}$.
We note that for spin rotation invariant situations, $J_{z}=J_{\parallel}$ and $J_{z}^{\prime}=J_{\parallel}^{\prime}$, $M=0$ and 
the effective Hamiltonian reduces to the time-averaged Hamiltonian.
This is expected since in this case $H_0$ and $H_1$ commute with each other.

\section{Appendix: Explicit form of first terms in the Magnus expansion for the interacting quantum spin chain \label{appendix-terms-magnus}}

As mentioned after Eq.~\eqref{eq:Kapitza-classical-1}, all the even order terms in the ME vanish for a time-symmetric driving~\cite{even-order1, even-order2}, i.e. $H(t)=H(T-t)$. Here, for completeness, we report the first not-vanishing terms in the ME for the protocol considered in the paper. First, it is convenient to introduce the shorthand notations $X\rightarrow  -\frac{i}{2\hbar}H_0T_0,\,\, Y\rightarrow -\frac{i}{\hbar}H_1T_1$ and $Z\rightarrow -\frac{i}{\hbar}H_{eff}T$ where $T=T_0+T_1$.
Then by the definition of evolution operator, the effective Hamiltonian is defined by $Z=\log\left[U(T)\right]=\log\left[e^{X}e^{Y}e^{X}\right]$.
Using a Mathematica code similar to the one described in Ref.~\cite{mathematicacode} we have explicitly computed the first few terms in the ME:
\begin{equation}
\begin{split}
&Z=\left(2X+Y\right)+\frac{1}{6}\left(-\left\{ XXY\right\} -\left\{ YXY\right\} \right)+\\
&\frac{1}{360}\biggl(7\left\{ XXXXY\right\} +\left\{ YYYXY\right\} +6\left\{ XYXXY\right\} \\
&+8\left\{ YXXXY\right\} +12\left\{ YYXXY\right\} -4\left\{ XYYXY\right\} \biggr)+...
\label{eq:5-commutators}
\end{split}
\end{equation}
where the symbol $\left\{ XYXXY\right\}$ is a short hand notation 
for the right nested commutators $\left[X,\left[Y,\left[X,\left[X,Y\right]\right]\right]\right]$.
We note that there is no unique way of writing $Z$ in terms of nested commutators due to the Jacoby identity of nested commutators.
For example, it is easy to show that $\left\{ YYXXY\right\} \equiv\left\{ YXYXY\right\}$. 
Substituting the definition for $X,Y,Z$ in Eq.~\eqref{eq:5-commutators} we obtain $H_{eff}=H_{eff}^{\left(1\right)}+H_{eff}^{\left(3\right)}+H_{eff}^{\left(5\right)}+...$ where:
\begin{equation}
\begin{array}{c}
H_{eff}^{\left(1\right)}=\frac{1}{T}\left(H_{0}T_{0}+H_{1}T_{1}\right)\\
H_{eff}^{\left(3\right)}=\frac{1}{6\hbar^{2}T}\left(\left(\frac{T_{0}}{2}\right)^{2}T_{1}\left\{ H_{0}H_{0}H_{1}\right\} +\left(\frac{T_{0}}{2}\right)T_{1}^{2}\left\{ H_{1}H_{0}H_{1}\right\} \right)\\
\begin{array}{c}
H_{eff}^{\left(5\right)}=\frac{1}{360\hbar^{4}T}\left(\left(\frac{T_{0}}{2}\right)^{4}T_{1}7\left\{ H_{0}H_{0}H_{0}H_{0}H_{1}\right\} +\left(\frac{T_{0}}{2}\right)T_{1}^{4}\left\{ H_{1}H_{1}H_{1}H_{0}H_{1}\right\} +\right.\\
\left(\frac{T_{0}}{2}\right)^{3}T_{1}^{2}\left(6\left\{ H_{0}H_{1}H_{0}H_{0}H_{1}\right\} +8\left\{ H_{1}H_{0}H_{0}H_{0}H_{1}\right\} \right)+\\
\left.\left(\frac{T_{0}}{2}\right)^{2}T_{1}^{3}\left(12\left\{ H_{1}H_{1}H_{0}H_{0}H_{1}\right\} -4\left\{ H_{0}H_{1}H_{1}H_{0}H_{1}\right\} \right)\right)
\end{array}
\end{array}
\end{equation}
These terms are different from the ones usually reported in the literature where the less symmetric protocol, $Z=\log\left[e^{X}e^{Y}\right]$, is considered.


\begin{thebibliography}{References}

\bibitem{thirring_02} W.~Thirring. {\em Quantum mathematical physics}, Springer, Berlin, 2002.

\bibitem{reif_09} F. Reif. {\em Fundamentals of statistical and thermal physics}, second ed., Waveland Press, Long Grove, 2009.

\bibitem{ap_ent} A. Polkovnikov. Microscopic diagonal entropy and its connection to basic thermodynamic relations. Annals Phys. {\bf 326} (2011) 486-499.

\bibitem{fermi_49} E. Fermi. On the Origin of the Cosmic Radiation. Phys. Rev. {\bf 75} (1949) 1169-1174.

\bibitem{LLX} E.~M.~Lifshitz, L.~Pitaevskii. {\em Physical Kinetics. Course of Theoretical Physics vol. 10}, third ed., Pergamon, 1981.

\bibitem{jarzynski_93} C. Jarzynski. Energy diffusion in a chaotic adiabatic billiard gas. Phys. Rev. E {\bf 48} (1993) 4340-4350.

\bibitem{srednicki_99} M.~Srednicki.  The approach to thermal equilibrium in quantized chaotic systems. J. Phys. A {\bf 32} (1999) 1163-1175.

\bibitem{rigol_08} M.~Rigol, V.~Dunjko, M.~Olshanii. Thermalization and its mechanism for generic isolated quantum systems. Nature {\bf 452} (2008) 854-858.

\bibitem{hanggi_11} M.~Campisi, P.~H\"anggi, P.~Talkner. Colloquium. Quantum Fluctuation Relations: Foundations and Applications. Rev. Mod. Phys. {\bf 83} (2011) 771-791.

\bibitem{lieberman_72} M. A. Lieberman and A. J. Lichtenberg. Stochastic and Adiabatic Behavior of Particles Accelerated by Periodic Forces. Phys. Rev. A {\bf 5} (1972) 1852-1866.

\bibitem{chirikov_81} B.V. Chirikov, F.M. Izrailev and D.L. Shepelyansky. Dynamical stochasticity in classical and quantum mechanics. Sov. Sci. Rev. C {\bf 2} (1981) 209.

\bibitem{fishman_82} S.~Fishman, D. R. Grempel, and R. E. Prange. Chaos, Quantum Recurrences, and Anderson Localization. Phys. Rev. Lett. {\bf 49} (1982) 509-512. 

\bibitem{Kapitza} P.L. Kapitza. Dynamic stability of a pendulum when its point of suspension vibrates. Soviet Phys. JETP {\bf21} (1951) 588.

\bibitem{Kapitza-math} H.W. Broer, I. Hoveijn, M. van Noort, C. Simo' and G. Vegter. The Parametrically Forced Pendulum: A Case Study in 1 1/2 Degree of Freedom. Journal of Dynamics and Differential Equations, \textbf{16} (2004) 897-947.

\bibitem{chapman} T.M. Hoang, C.S. Gerving, B.J. Land, M. Anquez, C.D. Hamley and M.S. Chapman. Dynamic stabilization of a quantum many-body system. arXiv:1209.4363v1 [cond-mat.quant-gas]

\bibitem{basko_06} D. M. Basko, I. L. Aleiner, and B. L. Altshuler. Metal-insulator transition in a weakly interacting many-electron system with localized single-particle states. Ann. Phys. {\bf 321} (2006) 1126-1205. 

\bibitem{pal_10} A. Pal and D. A. Huse. Many-body localization phase transition. Phys. Rev. B {\bf 82} (2010) 174411.

\bibitem{Boris}Kai Ji and B. V. Fine. Nonthermal Statistics in Isolated Quantum Spin Clusters after a Series of Perturbations. Phys. Rev. Lett. \textbf{107} (2011) 050401 

\bibitem{Ovadyahu} Z. Ovadyahu. Suppression of Inelastic Electron-Electron scattering in Anderson Insulators. Phys. Rev. Lett. {\bf 108} (2012) 156602

\bibitem{fishman-1} S. Rahav, I. Gilary and S. Fishman, Effective Hamiltonians for periodically driven systems, Phys. Rev. A \textbf{68} (2003) 013820 


\bibitem{fishman-2} S. Rahav, I. Gilary and S. Fishman, Time Independent Description of Rapidly Oscillating Potentials, Phys. Rev. Lett. \textbf{91} (2003) 110404 

\bibitem{Landau} L.D. Landau and E. M. Lifshitz. {\em Mechanics}, Pergamon, Oxford, 1976. 



\bibitem{Magnus}W. Magnus. On the exponential solution of differential equations for a linear operator. Commun. Pure Appl. Math. \textbf{7} (1954) 649-673. 

\bibitem{Magnus-report}S. Blanes, F.Casas. J. A. Oteo and J.Ros. The Magnus expansion and some of its applications. Physics Reports \textbf{470} (2009) 151-238.  

\bibitem{lindner_11} N.~H.~Lindner, G.~Refael, V.~Galitski. Floquet Topological Insulator in Semiconductor Quantum Wells. Nat. Phys. {\bf 7} (2011) 490-495. 

\bibitem{kitagawa_12} T.~Kitagawa, M.~A.~Broome, A.~Fedrizzi, M.~S.~Rudner, E.~Berg, I.~Kassal, A.~Aspuru-Guzik, E.~Demler, Andrew G. White. Observation of topologically protected bound states in a one dimensional photonic system. Nat. Commun. {\bf 3} (2012) 882

\bibitem{kardar} M. Kardar. {\em Statistical physics of particles}, Cambridge University Press, Cambridge, 2007.

\bibitem{even-order1} A. Iserles, S.P. Norsett, A.F. Rasmussen. Time symmetry and high-order Magnus methods. Appl. Numer. Math. {\bf39} (2001) 379-401. 

\bibitem{even-order2} H. Munthe-Kaas, B. Owren. Computations in a free Lie algebra. Phil. Trans. R. Soc. A 357 (1999) 957-981.

\bibitem{bunin_11} G.~Bunin, L.~D'Alessio, Y.~Kafri, A.~Polkovnikov. Universal energy fluctuations in thermally isolated driven systems. Nat. Phys {\bf 7} (2011) 913-917.

\bibitem{BCH} K. Goldberg. The formal power series for $log e^x e^y$. Duke Math. J. {\bf 23} (1956) 13-21.

\bibitem{heyl_12} M.~Heyl, A.~Polkovnikov, S.~Kehrein. Dynamical Quantum Phase Transitions in the Transverse Field Ising Model.  arXiv:1206.2505.

\bibitem{ap_phasespace} A. Polkovnikov. Phase space representation of quantum dynamics. Annals of Phys. {\bf325} (2010) 1790-1852. 

\bibitem{ikeda} T. N. Ikeda, unpublished

\bibitem{mathematicacode} Matthias W. Reinsch. A simple expression for the terms in the Baker-Campbell-Hausdorff series. arXiv:math-ph/9905012v2

\bibitem{Kehrein_1} M. Heyl and S. Kehrein. Nonequilibrium steady state in a periodically driven Kondo model. Phys. Rev. B {\bf 81} (2010) 144301

\bibitem{Kehrein_2} M. Heyl and S. Kehrein. Private Communications. 

%\bibitem{future} L. D'Alessio, K. Ji, B. Fine and A. Polkovnikov (in preparation)


% prosen PRL 107 060403 (2011)
% experimental results


\end{thebibliography}
\end{document}